\documentclass{article}%
\usepackage{amsfonts}
\usepackage{amsmath}
\usepackage{amssymb}
\usepackage{graphicx}%
\providecommand{\U}[1]{\protect\rule{.1in}{.1in}}
\newtheorem{theorem}{Theorem}
\newtheorem{acknowledgement}[theorem]{Acknowledgement}

\newtheorem{definition}[theorem]{Definition}

\begin{document}

\title{Chern-Simons and Born-Infeld gravity theories and Maxwell algebras type}
\author{P.K. Concha, D. M. Pe\~{n}afiel , E. K Rodriguez, \ P. Salgado\\Departamento de F\'{\i}sica, Universidad de Concepci\'{o}n \\Casilla 160-C, Concepci\'{o}n, Chile}
\maketitle

\begin{abstract}
Recently was shown that standard odd and even-dimensional General Relativity
can be obtained from a $\left(  2n+1\right)  $-dimensional Chern-Simons
Lagrangian invariant under the $B_{2n+1}$ algebra and from a $\left(
2n\right)  $-dimensional Born-Infeld Lagrangian invariant under a subalgebra
$\mathcal{L}^{B_{2n+1}}$ respectively.

Very Recently, it was shown that the generalized In\"{o}n\"{u}-Wigner
contraction of the generalized AdS-Maxwell algebras provides Maxwell algebras
types $\mathcal{M}_{m}$ which correspond to the so called  $B_{m}$ Lie algebras.

In this article we report on a simple model that suggests a mechanism by which
standard \ odd-dimensional General Relativity may emerge as a weak coupling
constant limit of a $\left(  2p+1\right)  $-dimensional Chern-Simons
Lagrangian invariant under the Maxwell algebra type $\mathcal{M}_{2m+1}$, if
and only if $m\geq p$. \ Similarly, we show that standard even-dimensional
General Relativity emerges as a weak coupling constant limit of a $\left(
2p\right)  $-dimensional Born-Infeld type Lagrangian invariant under a
subalgebra $\mathcal{L}^{\mathcal{M}_{\mathbf{2m}}}$ of the Maxwell algebra
type, if and only if $m\geq p$. \ It is shown that when $m<p$ this is not
possible for $\ $a $(2p+1)$-dimensional Chern-Simons Lagrangian \ invariant
under the $\mathcal{M}_{2m+1}$\ \ and for a $\left(  2p\right)  $-dimensional
Born-Infeld type Lagrangian invariant under $\mathcal{L}^{\mathcal{M}%
_{\mathbf{2m}}}$ algebra.

\end{abstract}

\section{\textbf{Introduction}}

The most general action for the metric satisfying the criteria of general
covariance and second-order field equations for $d>4$ is a polynomial of
degree $\left[  d/2\right]  $ in the curvature known as the Lanczos-Lovelock
gravity theory $\left(  LL\right)  $ \cite{lanc},\cite{lovel}. The $LL$
lagrangian in a $d$-dimensional Riemannian manifold can be defined as a linear
combination of the dimensional continuation of all the Euler classes of
dimension $2p<d$ \cite{zum},\cite{teit}:
\begin{equation}
S=\int\sum_{p=0}^{\left[  d/2\right]  }\alpha_{p}L^{(p)} \label{uno}%
\end{equation}
where $\alpha_{p}$ are arbitrary constants and
\begin{equation}
L_{p}=\varepsilon_{a_{1}a_{2}\cdot\cdot\cdot\cdot\cdot\cdot a_{d}}%
R^{a_{1}a_{2}}\cdot\cdot\cdot\cdot R^{a_{2p-1}a_{2p}}e^{a_{2p+1}}\cdot
\cdot\cdot\cdot e^{a_{d}} \label{dos}%
\end{equation}
with $R^{ab}=d\omega^{ab}+\omega_{c}^{a}\omega^{cb}.$ The expression
(\ref{uno}) can be used both for even and for odd dimensions.

The large number of dimensionful constants in the $LL$ theory $\alpha_{p},$
$p=0,1,\cdot\cdot\cdot,\left[  d/2\right]  ,$ which are not fixed from first
principles, contrast with the two constants of the Einstein-Hilbert action.

In ref. \cite{tron} it was found that these parameters can be fixed in terms
of the gravitational and the cosmological constants, and that the action in
odd dimensions can be formulated as a Chern-Simons theory of the $AdS$ group.

The closest one can get to a Chern-Simons theory in even dimensions is with
the so-called Born-Infeld theories \cite{tron} \cite{ban1}, \cite{ban2}%
,\cite{zanel}. The Born-Infeld lagrangian is obtained by a particular choice
of the parameters in the Lovelock series, so that the lagrangian is invariant
only under local Lorentz rotations in the same way as is the Einstein-Hilbert action.

If Chern-Simons theory is the appropriate odd-dimensional gauge theory and
if\ Born-Infeld theory is the appropriate even-dimensional theories\ to
provide a\ framework for the gravitational interaction, then these theories
must satisfy the correspondence principle, namely they must be related to
General Relativity. \ 

In Ref. \cite{salg1} was shown that the standard, odd-dimensional General
Relativity (without a cosmological constant) can be obtained from Chern-Simons
gravity theory for a certain Lie algebra $\mathfrak{B}$ and recently was found
that\ standard, even-dimensional General Relativity (without a cosmological
constant) emerges as a limit of a Born-Infeld theory invariant under a certain
subalgebra of the Lie algebra $\mathfrak{B}$ \cite{salg1a}.

Very recently was found that the so called $\mathfrak{B}_{m}$ Lie algebra of
Ref. \cite{salg1} correspond to\ Maxwell algebras type $\mathcal{M}_{m}%
$\ \cite{salg1c}. In fact, it was shown that the generalized
In\"{o}n\"{u}-Wigner contraction of the generalized AdS-Maxwell algebras
provides maxwell algebras types $M_{\mathbf{m}}$ which correspond to
$B_{\mathbf{m}}$ Lie algebra. \ These Maxwell algebras type $M_{m}$ algebras
can be obtained \ by $S$-expansion resonant reduced of the $AdS$ Lie algebra
when we use $S_{E}^{(N)}=\left\{  \lambda_{\alpha}\right\}  _{\alpha=0}^{N+1}$
as semigroup.

It is the purpose of this paper to show that standard odd General Relativity
emerges as a weak coupling constant limit of a $\left(  \mathbf{2p+1}\right)
$-dimensional Chern-Simons Lagrangian invariant under the $M_{\mathbf{2m+1}}$
algebra, if and only if $m\geq p$. \ Similarly, we show that standard even
General Relativity emerges as a weak coupling constant limit of a $\left(
\mathbf{2p}\right)  $-dimensional Born-Infeld type Lagrangian invariant under
the $L^{\mathcal{M}_{\mathbf{2m}}}$ algebra, if and only if $m\geq p$.  \ It
is shown that when $m<p$  this is not possible for $\ $a $(2p+1)$-dimensional
Chern-Simons Lagrangian \ invariant under the $M_{\mathbf{2m+1}}$\ \ and for a
$\left(  \mathbf{2p}\right)  $-dimensional Born-Infeld type Lagrangian
invariant under $L^{\mathcal{M}_{\mathbf{2m}}}$.

This paper is organized as follows: In Sec.~II we briefly review some aspect
of: $(i)$ Lovelock gravity theory, $(ii)$ the construction of the so called
$\mathcal{M}_{2n+1}$ algebra and $(iii)$ obtaining odd and even dimensional
general relativity from Chern-Simons gravity theory and from Born-Infeld
theory respectively.

In Section III it is shown that the odd-dimensional Einstein-Hilbert
Lagrangian can be obtained from a Chern-Simons Lagrangian\textbf{ }in
$(2p+1)$-dimensions invariant under the algebra $M_{\mathbf{2m+1}}$, if and
only if $m\geq p$. However, this is not possible for Chern-Simons Lagrangian
in $\left(  \mathbf{2p+1}\right)  $-dimension invariant under the
$M_{\mathbf{2m+1}}$ algebra when $m<p$.

In Section IV it is shown that the even-dimensional Einstein-Hilbert
Lagrangian can be obtained from a Born-Infeld type Lagrangian in\ $\left(
\mathbf{2p}\right)  $-dimensions\ invariant under the $L^{\mathcal{M}%
_{\mathbf{2m}}}$ subalgebra of the $M_{\mathbf{2m}}$ algebra , if and only if
$m\geq p$. However, this is not possible for Born-Infeld type Lagrangians in
$\left(  \mathbf{2p}\right)  $-dimensions invariant under the $L^{\mathcal{M}%
_{\mathbf{2m}}}$ subalgebra when $m<p.$

Sec.V concludes the work with a comment about possible developments. \ 

\section{\textbf{The Lovelock action, The }$\mathcal{M}_{2n+1}$%
\textbf{\ algebra and general relativity }}

In this section we shall review some aspects of higher dimensional gravity,
the construction of the so called Maxwell algebra types, and obtaining odd and
even dimensional general relativity from Chern-Simons gravity theory and from
Born-Infeld theory respectively. The main point of this section is to display
the differences between the invariances of Lovelock action when odd and even
dimensions are considered.

\subsection{\textbf{The Chern-Simons gravity}}

The Lovelock action is a polynomial of degree $\left[  d/2\right]  $ in
curvature, which can be written in terms of the Riemann curvature and the
vielbein $e^{a}$ in the form (\ref{uno}), (\ref{dos}). In first order
formalism the Lovelock action is regarded as a functional of the vielbein and
spin connection, and the corresponding field equations obtained by varying
with respect to $e^{a}$ and $\omega^{ab}$ read \cite{tron}:%

\begin{equation}
\varepsilon_{a}=\sum_{p=0}^{\left[  \left(  d-1\right)  /2\right]  }\alpha
_{p}(d-2p)\varepsilon_{a}^{p}=0;\text{ \ }\varepsilon_{ab}=\sum_{p=1}^{\left[
\left(  d-1\right)  /2\right]  }\alpha_{p}p(d-2p)\varepsilon_{ab}^{p}=0
\label{tres}%
\end{equation}
where
\begin{equation}
\varepsilon_{a}^{p}:=\varepsilon_{ab_{1}\cdot\cdot\cdot b_{d-1}}R^{b_{1}b_{2}%
}\cdot\cdot\cdot R^{b_{2p-1}b_{2p}}e^{b_{2p+1}}\cdot\cdot\cdot e^{b_{d-1}}
\label{4}%
\end{equation}%
\begin{equation}
\varepsilon_{ab}^{p}=\varepsilon_{aba_{3}\cdot\cdot\cdot a_{d}}R^{a_{3}a_{4}%
}\cdot\cdot\cdot R^{a_{2p-1}a_{2p}}T^{a_{2p+1}}e^{a_{2p+2}}\cdot\cdot
\cdot\cdot e^{a_{d}}. \label{5}%
\end{equation}
Here $T^{a}=de^{a}+\omega_{b}^{a}e^{b}$ is the torsion $2$-form. Using the
Bianchi identity one finds \cite{tron}
\begin{equation}
D\varepsilon_{a}=\sum_{p=1}^{\left[  \left(  d-1\right)  /2\right]  }%
\alpha_{p-1}(d-2p+2)(d-2p+1)e^{b}\varepsilon_{ba}^{p}. \label{siete}%
\end{equation}
Moreover
\begin{equation}
e^{b}\varepsilon_{ba}=\sum_{p=1}^{\left[  \left(  d-1\right)  /2\right]
}\alpha_{p}p(d-2p)e^{b}\varepsilon_{ba}^{p}. \label{ocho}%
\end{equation}

From (\ref{siete}) and (\ref{ocho}) one finds for $d=2n-1$%
\begin{equation}
\alpha_{p}=\alpha_{0}\frac{(2n-1)(2\gamma)^{p}}{(2n-2p-1)}\left(
\genfrac{}{}{0pt}{}{n-1}{p}%
\right)  ; \label{nueve}%
\end{equation}
with $\alpha_{0}=\frac{\kappa}{dl^{d-1}},$ $\gamma=-sign(\Lambda)\frac{l^{2}%
}{2},$ where for any dimensions $l$ is a length parameter related to the
cosmological constant by $\Lambda=\pm(d-1)(d-2)/2l^{2}$

With these coefficients, the Lovelock action is a Chern-Simons $\left(
2n-1\right)  $-form invariant not only under standard local Lorentz rotations
$\delta e^{a}=\kappa_{\text{ }b}^{a}e^{b},\quad\delta\omega^{ab}=-D\kappa
^{ab},$ but also under a local $AdS$ boost \cite{tron}.

\subsection{\textbf{Born-Infeld gravity}}

For $d=2n$ it is necessary to write equation (\ref{siete}) in the form
\cite{tron}
\begin{equation}
D\varepsilon_{a}=T^{\mathbf{b}}\sum_{p=1}^{\left[  n-1\right]  }2\alpha
_{p-1}(n-p+1)\mathcal{T}_{ab}^{p}-\sum_{p=1}^{\left[  n-1\right]  }%
4\alpha_{p-1}(n-p+1)(n-p)e^{b}\varepsilon_{ba}^{p} \label{diez}%
\end{equation}
with
\begin{equation}
\mathcal{T}_{ab}=\frac{\delta L}{\delta R^{ab}}=\sum_{p=1}^{\left[  \left(
d-1\right)  /2\right]  }\alpha_{p}p\mathcal{T}_{ab}^{p} \label{once}%
\end{equation}
where
\begin{equation}
\mathcal{T}_{ab}^{p}=\varepsilon_{aba_{3}\cdot\cdot\cdot\cdot a_{d}}%
R^{a_{3}a_{4}}\cdot\cdot\cdot\cdot\cdot\cdot R^{a_{2p-1}a_{2p}}T^{a_{2p+1}%
}e^{a_{2p+2}}\cdot\cdot\cdot e^{a_{d}}. \label{doce}%
\end{equation}

The comparison between (\ref{ocho}) and (\ref{diez}) leads to \cite{tron}
\begin{equation}
\alpha_{p}=\alpha_{0}(2\gamma)^{p}\binom np . \label{trece}%
\end{equation}

With these coefficients the $LL$ lagrangian takes the form \cite{tron}%
\begin{equation}
L=\frac{\kappa}{2n}\varepsilon_{a_{1}a_{2}\cdot\cdot\cdot\cdot\cdot\cdot
a_{d}}\bar{R}^{a_{1}a_{2}}\cdot\cdot\cdot\cdot\cdot\cdot\bar{R}^{a_{d-1}a_{d}}
\label{13}%
\end{equation}
which is the Pfaffian of the 2-form $\bar{R}^{ab}=R^{ab}+\frac{1}{l^{2}}%
e^{a}e^{b}$ and can be formally written as the Born-Infeld like form
\cite{tron},\cite{zanel} . The corresponding action, known as Born-Infeld
action is invariant only under local Lorentz rotations.

The corresponding Born-Infeld action is given by \cite{tron}, \cite{zanel}%

\begin{equation}
S=\int\sum_{p=0}^{\left[  d/2\right]  }\frac{\kappa}{2n}\binom{n}{p}%
l^{2p-d+1}\varepsilon_{a_{1}\cdot\cdot\cdot\cdot a_{d}}R^{a_{1}a_{2}}%
\cdot\cdot\cdot R^{a_{2p-1}a_{2p}}e^{a_{2p+1}}\cdot\cdot\cdot e^{a_{d}}.
\label{14}%
\end{equation}
where $e^{a}$ corresponds to the 1-form \emph{vielbein}, and $R^{ab}%
=\mathrm{d}\omega^{ab}+\omega_{\text{ \ \ }c}^{a}\omega^{cb}$ to the Riemann
curvature in the first order formalism.

The action (\ref{14}) is off-shell invariant under the Lorentz-Lie algebra
$\mathrm{SO}\left(  2n-1,1\right)  ,$\ whose generators $\boldsymbol{\tilde
{J}}_{ab}$ of Lorentz transformations satisfy the commutation relationships%
\[
\left[  \boldsymbol{\tilde{J}}_{ab},\boldsymbol{\tilde{J}}_{cd}\right]
=\eta_{cb}\boldsymbol{\tilde{J}}_{ad}-\eta_{ca}\boldsymbol{\tilde{J}}%
_{bd}+\eta_{db}\boldsymbol{\tilde{J}}_{ca}-\eta_{da}\boldsymbol{\tilde{J}%
}_{cb}%
\]

The Levi-Civita symbol $\varepsilon_{a_{1}...a_{2n}}$ in (\ref{14}) should be
regarded as the only non-vanishing component of the symmetric, $\mathrm{SO}%
\left(  2n-1,1\right)  ,$ invariant tensor of rank $n,$ namely
\begin{equation}
\left\langle \boldsymbol{\tilde{J}}_{a_{1}a_{2}}\cdots\boldsymbol{\tilde{J}%
}_{a_{2n-1}a_{2n}}\right\rangle =\frac{2^{n}}{n}\epsilon_{a_{1}\cdots a_{2n}}.
\end{equation}

In order to interpret the gauge field as the vielbein, one is \textsl{forced}
to introduce a length scale $l$ in the theory. To see why this happens,
consider the following argument: Given that $(i)$ the exterior derivative
operator $\mathrm{d}=\mathrm{d}x^{\mu}\partial_{\mu}$ is dimensionless, and
$(ii)$ one always chooses Lie algebra generators $T_{A}$ to be dimensionless
as well, the one-form connection fields $\boldsymbol{A}=A_{\text{ \ }\mu}%
^{A}\boldsymbol{T}_{A}\mathrm{d}x^{\mu}$ must also be dimensionless. However,
the vielbein $e^{a}=e_{\text{ \ }\mu}^{a}\mathrm{d}x^{\mu}$ must have
dimensions of length if it is to be related to the spacetime metric $g_{\mu
\nu}$ through the usual equation $g_{\mu\nu}=e_{\text{ \ }\mu}^{a}e_{\text{
\ }\nu}^{b}\eta_{ab}.$ This means that the \textquotedblleft
true\textquotedblright\ gauge field must be of the form $e^{a}/l$, with $l$ a
length parameter.

Therefore, following Refs.~\cite{Cha89}, \cite{Cha90}, the one-form gauge
field $\boldsymbol{A}$ of the Chern--Simons theory is given in this case by%
\begin{equation}
\boldsymbol{A}=\frac{1}{l}e^{a}\boldsymbol{\tilde{P}}_{a}+\frac{1}{2}%
\omega^{ab}\boldsymbol{\tilde{J}}_{ab}. \label{ehcs0}%
\end{equation}

It is important to notice that once the length scale $l$ is brought into the
Born-Infeld theory, the lagrangian splits into several sectors, each one of
them proportional to a different power of $l$, as we can see directly in
eq.~(\ref{14}).

\subsection{\textbf{The Maxwell algebra type}}

\subsubsection{\textbf{The S-expansion procedure}}

In this subsection we shall review the main aspects of the $S$-expansion
procedure and their properties introduced in Ref. \cite{salg2}.

Let $S=\left\{  \lambda_{\alpha}\right\}  $ be an abelian semigroup with
2-selector $K_{\alpha\beta}^{\ \ \ \gamma}$ defined by
\begin{equation}
K_{\alpha\beta}^{\ \ \ \gamma}=\left\{
\begin{array}
[c]{cc}%
1 & \ \ \ \ \ \ \ \ \ \ \mbox{when}\ \ \lambda_{\alpha}\lambda_{\beta}%
=\lambda_{\gamma}\\
0 & \mbox{otherwise},
\end{array}
\right.
\end{equation}
and $\mathfrak{g}$ a Lie (super)algebra with basis $\left\{  \mathbf{T}%
_{A}\right\}  $ and structure constant $C_{AB}^{\ \ \ C}$,
\begin{equation}
\left[  \mathbf{T}_{A},\mathbf{T}_{B}\right]  =C_{AB}^{\ \ \ C}\mathbf{T}_{C}.
\end{equation}
Then it may be shown that the product $\mathfrak{G}=S\times\mathfrak{g}$ is
also a Lie (super)algebra with structure constants $C_{(A,\alpha)(B,\beta
)}^{\ \ \ \ \ \ \ \ \ \ \ \ (C,\gamma)}=K_{\alpha\beta}^{\ \ \gamma}%
C_{AB}^{\ \ \ \ C}$,
\begin{equation}
\left[  \mathbf{T}_{(A,\alpha)},\mathbf{T}_{(B,\beta)}\right]  =C_{(A,\alpha
)(B,\beta)}^{\ \ \ \ \ \ \ \ \ \ \ \ (C,\gamma)}\mathbf{T}_{(C,\gamma)}.
\end{equation}
The proof is direct and may be found in Ref. \cite{salg2}.

\begin{definition}
Let $S$ be an abelian semigroup and $\mathfrak{g}$ a Lie algebra. The Lie
algebra $\mathfrak{G}$ defined by $\mathfrak{G}=S\times\mathfrak{g}$ is called
$S$-Expanded algebra of $\mathfrak{g}$.
\end{definition}

When the semigroup has a zero element $0_{S}\in S$, it plays a somewhat
peculiar role in the $S$-expanded algebra. The above considerations motivate
the following definition:

\begin{definition}
Let $S$ be an abelian semigroup with a zero element $0_{S}\in S$, and let
$\mathfrak{G}=S\times\mathfrak{g}$ be an $S$-expanded algebra. The algebra
obtained by imposing the condition $0_{S}\mathbf{T}_{A}=0$ on $\mathfrak{G}$
(or a subalgebra of it) is called $0_{S}$-reduced algebra of $\mathfrak{G}$
(or of the subalgebra).
\end{definition}

An $S$-expanded algebra has a fairly simple structure. Interestingly, there
are at least two ways of extracting smaller algebras from $S\times
\mathfrak{g}$. The first one gives rise to a \textit{resonant subalgebra},
while the second produces reduced algebras. \ In particular, a resonant
subalgebra can be obtained as follow.

Let $g=%
{\textstyle\bigoplus_{p\in I}}
V_{p}$ be a decomposition of $g$ in subspaces $V_{p}$, where $I$ is a set of
indices. \ For each $p,q\in I$ it is always possible to define $i_{\left(
p,q\right)  }\subset I$ such that%
\begin{equation}
\left[  V_{p},V_{q}\right]  \subset%
{\textstyle\bigoplus\limits_{r\in i_{\left(  p,q\right)  }}}
V_{r},\label{eq33}%
\end{equation}
\textbf{ }Now, let $S=%
{\textstyle\bigcup_{p\in I}}
S_{p}$ be a subset decomposition of the abelian semigroup $S$ such that%
\begin{equation}
S_{p}\cdot S_{q}\subset%
{\textstyle\bigcup_{r\in i_{\left(  p,q\right)  }}}
S_{p}.\label{eq34}%
\end{equation}
When such subset decomposition $S=%
{\textstyle\bigcup_{p\in I}}
S_{p}$ exists, then we say that this decomposition is in resonance with the
subspace decomposition of $g,$ $g=%
{\textstyle\bigoplus_{p\in I}}
V_{p}$.

The resonant subset decomposition is crucial in order to systematically
extract subalgebras from the $S$-expanded algebra $G=S\times g$, as is proven
in the following\medskip

Theorem IV.2 of Ref. \cite{salg2}: Let $g=%
{\textstyle\bigoplus_{p\in I}}
V_{p}$ be a subspace decomposition of $g$, with a structure described by eq.
$\left(  \ref{eq33}\right)  ,$ and let $S=%
{\textstyle\bigcup_{p\in I}}
S_{p}$ be a resonant subset decomposition of the abelian semigroup $S$, with
the structure given in eq. $\left(  \ref{eq34}\right)  $. Define the subspaces
of $G=S\times g$,%
\begin{equation}
W_{p}=S_{p}\times V_{p},\text{ \ }p\in I.
\end{equation}
Then,%
\begin{equation}
\mathfrak{G}_{R}=%
{\textstyle\bigoplus_{p\in I}}
W_{p}%
\end{equation}
is a subalgebra of $G=S\times g$.

Proof: \ the proof may be found in Ref. \cite{salg2}.

\begin{definition}
The algebra $G_{R}=%
{\textstyle\bigoplus_{p\in I}}
W_{p}$ obtained is called a Resonant Subalgebra of the $S$-expanded algebra
$G=S\times g$.
\end{definition}

A useful property of the $S$-expansion procedure is that it provides us with
an invariant tensor for the $S$-expanded algebra $\mathfrak{G}=S\times
\mathfrak{g}$ in terms of an invariant tensor for $\mathfrak{g}$. As shown in
Ref. \cite{salg2} the theorem VII.2 provide a general expression for an
invariant tensor for a $0_{S}$-reduced algebra.

\textbf{Theorem VII.2 of Ref. \cite{salg2}:} \ Let $S$ be an abelian semigroup
with nonzero elements $\lambda_{i}$, $i=0,\cdots,N$ and $\lambda_{N+1}=0_{S}$.
Let $\mathfrak{g}$ be a Lie (super)algebra of basis $\left\{  \mathbf{T}%
_{A}\right\}  $, and let $\langle\mathbf{T}_{A_{n}}\cdots\mathbf{T}_{A_{n}%
}\rangle$ be an invariant tensor for $\mathfrak{g}$. The expression
\begin{equation}
\langle\mathbf{T}_{(A_{1},i_{1})}\cdots\mathbf{T}_{(A_{n},i_{n})}%
\rangle=\alpha_{j}K_{i_{a}\cdots i_{n}}^{\ \ \ \ \ j}\langle\mathbf{T}_{A_{1}%
}\cdots\mathbf{T}_{A_{n}}\rangle
\end{equation}

where $\alpha_{j}$ are arbitrary constants, corresponds to an invariant tensor
for the $0_{S}$-reduced algebra obtained from $\mathfrak{G}=S\times
\mathfrak{g}$.

\textbf{Proof:} \ the proof may be found in section $4.5$ of Ref. \cite{salg2}.

\subsubsection{\textbf{S-expansion of }$SO\left(  2n,2\right)  $\textbf{
algebra}}

Let us consider the $S$-expansion of the Lie algebra $\mathrm{SO}\left(
2n,2\right)  $ using the Abelian semigroup $S_{\mathrm{E}}^{\left(
2n-1\right)  }=\left\{  \lambda_{0},\lambda_{1},\lambda_{2},\lambda
_{3},\lambda_{4},\lambda_{5},\lambda_{6},\cdot\cdot\cdot,\lambda_{2n}\right\}
$ defined by the product%
\begin{equation}
\lambda_{\alpha}\lambda_{\beta}=\left\{
\begin{array}
[c]{c}%
\lambda_{\alpha+\beta},\text{ \ \ when }\alpha+\beta\leq2n\\
\lambda_{2n},\text{ when \ }\alpha+\beta>2n\text{\ \ \ \ \ \ \ \ \ \ \ \ \ }%
\end{array}
\right.
\end{equation}

The $\lambda_{\alpha}$ elements are dimensionless, and can be represented by
the set of $2n\times2n$ matrices $\left[  \lambda_{\alpha}\right]  _{\text{
}j}^{i}=\delta_{\text{ \ }j+\alpha}^{i}$, where $i,j=1,\cdot\cdot\cdot,2n-1$,
$\alpha=0,\cdot\cdot\cdot\cdot,2n$, and $\delta$ stands for the Kronecker
delta \cite{salg1}.

After extracting a resonant subalgebra and perfoming its $0_{S}(=\lambda
_{2n})$-reduction, one finds a new Lie algebra the so called Maxwell algebra
type $\mathcal{M}_{2n+1}$, which in Ref. \cite{salg1} was called
$\mathfrak{B}_{2n+1}$ algebra$\mathfrak{,}$ whose generators%

\begin{align}
J_{\left(  ab,2k\right)  }  &  =\lambda_{2k}\otimes\tilde{J}_{ab},\\
P_{\left(  a,2k+1\right)  }  &  =\lambda_{2k+1}\otimes\tilde{P}_{a},
\end{align}
with $k=0,\cdot\cdot\cdot\cdot,n-1,$ satisfy the commutation relationships
\cite{salg1}%

\begin{align}
\left[  P_{a},P_{b}\right]   &  =Z_{ab}^{\left(  1\right)  },\text{
\ \ \ \ }\left[  J_{ab},P_{c}\right]  =\eta_{bc}P_{a}-\eta_{ac}P_{b}%
\label{1'}\\
\left[  J_{ab,}J_{cd}\right]   &  =\eta_{cb}J_{ad}-\eta_{ca}J_{bd}+\eta
_{db}J_{ca}-\eta_{da}J_{cb}\label{1}\\
\left[  J_{ab},Z_{c}^{\left(  i\right)  }\right]   &  =\eta_{bc}Z_{a}^{\left(
i\right)  }-\eta_{ac}Z_{b}^{\left(  i\right)  },\\
\left[  Z_{ab}^{\left(  i\right)  },P_{c}\right]   &  =\eta_{bc}Z_{a}^{\left(
i\right)  }-\eta_{ac}Z_{b}^{\left(  i\right)  },\\
\left[  Z_{ab}^{\left(  i\right)  },Z_{c}^{\left(  j\right)  }\right]   &
=\eta_{bc}Z_{a}^{\left(  i+j\right)  }-\eta_{ac}Z_{b}^{\left(  i+j\right)  }\\
\left[  J_{ab,}Z_{cd}^{\left(  i\right)  }\right]   &  =\eta_{cb}%
Z_{ad}^{\left(  i\right)  }-\eta_{ca}Z_{bd}^{\left(  i\right)  }+\eta
_{db}Z_{ca}^{\left(  i\right)  }-\eta_{da}Z_{cb}^{\left(  i\right)  }%
\label{2}\\
\left[  Z_{ab,}^{\left(  i\right)  }Z_{cd}^{\left(  j\right)  }\right]   &
=\eta_{cb}Z_{ad}^{\left(  i+j\right)  }-\eta_{ca}Z_{bd}^{\left(  i+j\right)
}+\eta_{db}Z_{ca}^{\left(  i+j\right)  }-\eta_{da}Z_{cb}^{\left(  i+j\right)
}\label{3}\\
\left[  P_{a},Z_{c}^{\left(  i\right)  }\right]   &  =Z_{ab}^{\left(
i+1\right)  },\text{ \ \ \ \ }\left[  Z_{a}^{\left(  i\right)  }%
,Z_{c}^{\left(  j\right)  }\right]  =Z_{ab}^{\left(  i+j+1\right)
}.\label{3'}%
\end{align}
and where we have defined%
\begin{align}
J_{ab} &  =J_{\left(  ab,0\right)  }=\lambda_{0}\otimes\tilde{J}_{ab},\\
P_{a} &  =P_{\left(  a,1\right)  }=\lambda_{1}\otimes\tilde{P}_{a},\\
Z_{ab}^{(i)} &  =J_{\left(  ab,2i\right)  }=\lambda_{2i}\otimes\tilde{J}%
_{ab},\\
Z_{a}^{\left(  i\right)  } &  =P_{\left(  a,2i+1\right)  }=\lambda
_{2i+1}\otimes\tilde{P}_{a},
\end{align}
\textbf{with }$i=1,...,n-1.$

We note that commutation relations (\ref{1}), (\ref{2}), (\ref{3}) form a
subalgebra of the $\mathcal{M}_{2n+1}$ algebra which we will denote as
$\mathfrak{L}^{\mathcal{M}_{2n+1}}.$ This subalgebra can be obtained from
$S$-expansion of the Lorentz-Lie algebra using as a semigroup the
sub-semigroup $S_{0}^{\left(  2n-1\right)  }=\left\{  \lambda_{0},\lambda
_{2},\lambda_{4},\lambda_{6},\cdot\cdot\cdot,\lambda_{2n}\right\}  $\ of
semigroup $S_{E}^{\left(  2n-1\right)  }=\left\{  \lambda_{0},\lambda
_{1},\lambda_{2},\lambda_{3},\lambda_{4},\lambda_{5},\lambda_{6},\cdot
\cdot\cdot,\lambda_{2n}\right\}  .$ After extracting a resonant subalgebra and
perfoming its $0_{S}(=\lambda_{2n})$-reduction, one finds the $\mathfrak{L}%
^{\mathcal{M}_{2n+1}}$ algebra$,$ which is a subalgebra of the $\mathcal{M}%
_{2n+1}$ algebra$\mathfrak{,}$ whose generators $J_{ab}=\lambda_{0}%
\boldsymbol{\tilde{J}}_{ab},$ $Z_{ab}^{(1)}=\lambda_{2}\boldsymbol{\tilde{J}%
}_{ab},$ $Z_{ab}^{(2)}=\lambda_{4}\boldsymbol{\tilde{J}}_{ab},\cdot\cdot
\cdot,Z_{ab}^{(n)}=\lambda_{2n}\boldsymbol{\tilde{J}}_{ab}$ satisfy the
commutation relationships (\ref{1}), (\ref{2}), (\ref{3}).

\subsection{\textbf{General Relativity}}

\subsubsection{\textbf{Odd-dimensional general relativity}}

In Ref. \cite{salg1}, it was shown that the standard, odd-dimensional general
relativity (without a cosmological constant) can be obtained from Chern-Simons
gravity theory for the algebra $\mathcal{M}_{2n+1}$. \ The Chern-Simons
Lagrangian is built from a $\mathcal{M}_{2n+1}$-valued, one-form gauge
connection $A$ which depends on a scale parameter $l$ which can be interpreted
as a coupling constant that characterizes diferent regimes whithin the theory.
\ The field content induced by $\mathcal{M}_{2n+1}$ includes the vielbein
$e^{a}$, the spin connection $\omega^{ab}$, and extra bosonic fields
$h^{a(i)}$ and $k^{ab(j)}.$ The odd-dimensional Chern-Simons Lagrangian
invariant under the $\mathcal{M}_{2n+1}$ algebra is given by \cite{salg1}%

\begin{align}
L_{CS}^{\mathfrak{(}2n+1\mathfrak{)}}  &  =\sum_{k=1}^{n}l^{2k-2}c_{k}%
\alpha_{j}\delta_{i_{1}+\cdots+i_{n+1}}^{j}\delta_{p_{1}+q_{1}}^{i_{k+1}%
}\cdots\delta_{p_{n-k}+q_{n-k}}^{i_{n}}\varepsilon_{a_{1}\cdots a_{2n+1}%
}\nonumber\\
&  \times R^{\left(  a_{1}a_{2},i_{1}\right)  }\cdot\cdot\cdot R^{\left(
a_{2k-1}a_{2k},i_{k}\right)  }e^{\left(  a_{2k+1},p_{1}\right)  }e^{\left(
a_{2k+2},q_{1}\right)  }\cdot\cdot\cdot\nonumber\\
&  \cdot\cdot\cdot e^{\left(  a_{2n-1},p_{n-k}\right)  }e^{\left(
a_{2n},q_{n-k}\right)  }e^{\left(  a_{2n+1},i_{n+1}\right)  }. \label{7}%
\end{align}
where
\[
c_{k}=\frac{1}{2(n-k)+1}\left(
\begin{array}
[c]{c}%
n\\
k
\end{array}
\right)
\]%
\[
R^{\left(  ab,2k\right)  }=d\omega^{(ab,2k)}+\eta_{cd}\omega^{(ac,2i)}%
\omega^{(db,2j)}\delta_{i+j}^{k}%
\]

In the $l\longrightarrow0$ limit, the only nonzero term in (\ref{7})
corresponds to the case $k=1$, whose only non-vanishing component occurs for
$p=q_{1}=\cdot\cdot\cdot\cdot=q_{2n-1}=0$ and is proportional to the
odd-dimensional Einstein-Hilbert Lagrangian \cite{salg1}%
\begin{equation}
\left.  L_{CS}^{\mathfrak{(}2n+1\mathfrak{)}}\right\vert _{l=0}=\frac
{n\alpha_{2n-1}}{2n-1}\varepsilon_{a_{1}\cdots a_{2n+1}}R^{a_{1}a_{2}}%
e^{a_{3}}\cdot\cdot\cdot\cdot e^{a_{2n+1}}%
\end{equation}

\subsubsection{\textbf{Even-dimensional general relativity}}

In Ref. \cite{salg1a}, it was recently shown that\ standard, even-dimensional
General Relativity (without a cosmological constant) emerges as a limit of a
Born-Infeld theory invariant under the subalgebra $\mathfrak{L}^{\mathcal{M}%
_{2n+1}}$ of the Lie algebra $\mathcal{M}_{2n+1}$.

The Born-Infeld Lagrangian is built from the two-form curvature $S_{0}%
^{\left(  2n-1\right)  }$-expanded\textbf{ }%
\begin{equation}
F=\sum_{k=0}^{n-1}\frac{1}{2}F^{\left(  ab,2k\right)  }J_{\left(
ab,2k\right)  },
\end{equation}
\textbf{where}%
\begin{align}
F^{\left(  ab,2k\right)  } &  =d\omega^{\left(  ab,2k\right)  }+\eta
_{cd}\omega^{\left(  ac,2i\right)  }\omega^{\left(  db,2j\right)  }%
\delta_{i+j}^{k}\nonumber\\
&  +\frac{1}{l^{2}}e^{\left(  a,2i+1\right)  }e^{\left(  b,2j+1\right)
}\delta_{i+j+1}^{k}%
\end{align}
which depends on a scale parameter $l$\ which can be interpreted as a coupling
constant that characterizes different regimes within the theory. The field
content induced by $L^{\mathcal{M}_{2n+1}}$\ includes the vielbein $e^{a}$,
the spin connection $\omega^{ab}$\ and extra bosonic fields $h^{a\left(
i\right)  }=e^{\left(  a,2i+1\right)  }$ and $k^{ab\left(  i\right)  }%
=\omega^{\left(  ab,2i\right)  },$ with $i=1,...,n-1$. \ The even-dimensional
Born-Infeld Gravity Lagrangian invariant under the $L^{\mathcal{M}_{2n+1}}$
algebra is given by \cite{salg1a}%

\begin{align}
L_{BI\text{ \ }(2n)}^{\mathfrak{L}^{\mathcal{M}}}  &  =\sum_{k=1}^{n}%
l^{2k-2}\frac{1}{2n}\binom{n}{k}\alpha_{j}\delta_{i_{1}+\cdots+i_{n}}%
^{j}\delta_{p_{1}+q_{1}}^{i_{k+1}}\cdots\delta_{p_{n-k}+q_{n-k}}^{i_{n}%
}\nonumber\\
&  \varepsilon_{a_{1}\cdots a_{2n}}R^{\left(  a_{1}a_{2},i_{1}\right)  }\cdots
R^{\left(  a_{2k-1}a_{2k},i_{k}\right)  }e^{\left(  a_{2k+1},p_{1}\right)
}\nonumber\\
&  e^{\left(  a_{2k+2},q_{1}\right)  }\cdots e^{\left(  a_{2n-1}%
,p_{n-k}\right)  }e^{\left(  a_{2n},q_{n-k}\right)  }.
\end{align}
where we can see that in the limit $l=0$ the only nonzero term corresponds to
the case $k=1$, \ whose only nonzero component (corresponding to the case
$p=q_{1}=\cdots=q_{2n-2}=0$) \cite{salg1a} is proportional to the
even-dimensional Einstein-Hilbert Lagrangian%

\begin{align}
\left.  L_{BI\text{ \ }(2n)}^{\mathfrak{L}^{\mathcal{M}}}\right\vert _{l=0}
&  =\frac{1}{2}\alpha_{2n-2}\varepsilon_{a_{1}\cdots a_{2n}}R^{\left(
a_{1}a_{2},0\right)  }e^{\left(  a_{3},1\right)  }\cdots e^{\left(
a_{2n},1\right)  }\nonumber\\
&  =\frac{1}{2}\alpha_{2n-2}\varepsilon_{a_{1}\cdots a_{2n}}R^{a_{1}a_{2}%
}e^{a_{3}}\cdots e^{a_{2n}}.
\end{align}

\section{\textbf{Chern-Simons Lagrangians invariant under the Maxwell algebra
type }}

In this section it is shown that the Einstein-Hilbert Lagrangian for
odd-dimensions can be obtained from a Chern-Simons Lagrangian in $\left(
2p+1\right)  $ dimensions invariant under the $M_{2m+1}$ algebra , if and only
if $m\geq p$. However, this is not possible when $m<p$ for Chern-Simons
Lagrangians in $\left(  2p+1\right)  $-dimensions invariant under the
$M_{2m+1}$ algebra .

The $1$-form gauge connection $A$ $\mathcal{M}_{2n+1}$-valued, is given by%
\begin{equation}
A=\sum_{k=0}^{n-1}\left[  \frac{1}{2}\omega^{\left(  ab,2k\right)  }J_{\left(
ab,2k\right)  }+\frac{1}{l}e^{\left(  a,2k+1\right)  }P_{\left(
a,2k+1\right)  }\right]  , \label{1''}%
\end{equation}
and the $2$-form curvature $F=dA+A^{2}$ is
\begin{equation}
F=\sum_{k=0}^{n-1}\left[  \frac{1}{2}F^{\left(  ab,2k\right)  }J_{\left(
ab,2k\right)  }+\frac{1}{l}F^{\left(  a,2k+1\right)  }P_{\left(
a,2k+1\right)  }\right]  , \label{2''}%
\end{equation}
where%
\begin{align}
F^{\left(  ab,2k\right)  }  &  =d\omega^{\left(  ab,2k\right)  }+\eta
_{cd}\omega^{\left(  ac,2i\right)  }\omega^{\left(  db,2j\right)  }%
\delta_{i+j}^{k}+\frac{1}{l^{2}}e^{\left(  a,2i+1\right)  }e^{\left(
b,2j+1\right)  }\delta_{i+j+1}^{k},\\
F^{\left(  a,2k+1\right)  }  &  =de^{\left(  a,2k+1\right)  }+\eta_{bc}%
\omega^{\left(  ab,2i\right)  }e^{\left(  c,2j\right)  }\delta_{i+j}^{k}.
\end{align}

It is interesting to note that the Maxwell algebra type $M_{2m+1}$ can be used
to construct different odd-dimensional Chern-Simons Lagrangians. \ For
example, if we consider a $S_{E}^{\left(  3\right)  }$-expansion of the $AdS$
algebra SO$\left(  4,2\right)  $ and after extracting a resonant subalgebra
and performing its $0_{S}$-reduction, one finds $M_{5}$ algebra in $D=5$
dimensions. \ On the other hand, if we consider a $S_{E}^{\left(  3\right)  }%
$-expansion of the $AdS$ algebra $SO\left(  6,2\right)  $ and after extracting
a resonant subalgebra and performing its $0_{S}$-reduction, one fnds $M_{5}$
algebra in $D=7\,\ $dimensions. \ In this way, the CS Lagrangians $L_{CS\text{
}\left(  5\right)  }^{\mathcal{M}_{5}}$ and $L_{CS\text{ }\left(  7\right)
}^{\mathcal{M}_{5}}$ are invariant under the same $M_{5}$ algebra, however the
indices of the generators $T_{a}$ runs over $5$ and $7$ values, respectively.

These considerations allow the construction of gravitational theories in every
odd-dimension. \ Nevertheless, as discussed below, only in some dimensions it
is possible to obtain General Relativity as a weak coupling constant limit of
a Chern-Simons theory.

\subsection{$(2+1)$\textbf{-dimensional Chern-Simons Lagrangians invariant
under }$\mathcal{M}_{7}$-\textbf{algebra}}

Before considering the Chern-Simons Lagrangian $\left(  2n+1\right)
$-dimensional, we study the case of the $\mathcal{M}_{7}$ algebra. \ The
$\mathcal{M}_{7}$-algebra can be found by $S$-expansion of the $AdS$ algebra
using as semigroup $S_{E}^{\left(  5\right)  }$. In fact, after extracting a
resonant subalgebra and performing the $0_{S}$ reduction, one finds the
$\mathcal{M}_{7}$-algebra whose generators satisfy the following commutation relations%

\begin{align}
\left[  P_{a},P_{b}\right]   &  =Z_{ab}^{\left(  1\right)  },\text{
\ \ \ \ }\left[  J_{ab},P_{c}\right]  =\eta_{bc}P_{a}-\eta_{ac}P_{b}\\
\left[  J_{ab,}J_{cd}\right]   &  =\eta_{cb}J_{ad}-\eta_{ca}J_{bd}+\eta
_{db}J_{ca}-\eta_{da}J_{cb}\\
\left[  J_{ab},Z_{c}^{\left(  1\right)  }\right]   &  =\eta_{bc}Z_{a}^{\left(
1\right)  }-\eta_{ac}Z_{b}^{\left(  1\right)  },\text{ \ \ \ \ }\left[
J_{ab},Z_{c}^{\left(  2\right)  }\right]  =\eta_{bc}Z_{a}^{\left(  2\right)
}-\eta_{ac}Z_{b}^{\left(  2\right)  }\\
\left[  Z_{ab}^{\left(  1\right)  },P_{c}\right]   &  =\eta_{bc}Z_{a}^{\left(
1\right)  }-\eta_{ac}Z_{b}^{\left(  1\right)  },\text{ \ \ \ \ }\left[
Z_{ab}^{\left(  2\right)  },P_{c}\right]  =\eta_{bc}Z_{a}^{\left(  2\right)
}-\eta_{ac}Z_{b}^{\left(  2\right)  }\\
\left[  Z_{ab}^{\left(  1\right)  },Z_{c}^{\left(  1\right)  }\right]   &
=\eta_{bc}Z_{a}^{\left(  2\right)  }-\eta_{ac}Z_{b}^{\left(  2\right)
},\text{ \ \ \ \ }\left[  P_{a},Z_{c}^{\left(  1\right)  }\right]
=Z_{ab}^{\left(  2\right)  }.\\
\left[  J_{ab,}Z_{cd}^{\left(  1\right)  }\right]   &  =\eta_{cb}%
Z_{ad}^{\left(  1\right)  }-\eta_{ca}Z_{bd}^{\left(  1\right)  }+\eta
_{db}Z_{ca}^{\left(  1\right)  }-\eta_{da}Z_{cb}^{\left(  1\right)  }\\
\left[  J_{ab,}Z_{cd}^{\left(  2\right)  }\right]   &  =\eta_{cb}%
Z_{ad}^{\left(  2\right)  }-\eta_{ca}Z_{bd}^{\left(  2\right)  }+\eta
_{db}Z_{ca}^{\left(  2\right)  }-\eta_{da}Z_{cb}^{\left(  2\right)  }\\
\left[  Z_{ab,}^{\left(  1\right)  }Z_{cd}^{\left(  1\right)  }\right]   &
=\eta_{cb}Z_{ad}^{\left(  2\right)  }-\eta_{ca}Z_{bd}^{\left(  2\right)
}+\eta_{db}Z_{ca}^{\left(  2\right)  }-\eta_{da}Z_{cb}^{\left(  2\right)  }\\
\left[  Z_{ab}^{\left(  2\right)  },Z_{c}^{\left(  1\right)  }\right]   &
=\left[  Z_{ab}^{\left(  2\right)  },Z_{c}^{\left(  2\right)  }\right]
=\left[  Z_{ab}^{\left(  1\right)  },Z_{c}^{\left(  2\right)  }\right]  =0,\\
\left[  Z_{ab,}^{\left(  2\right)  }Z_{cd}^{\left(  2\right)  }\right]   &
=\left[  Z_{ab,}^{\left(  1\right)  }Z_{cd}^{\left(  2\right)  }\right]
=\left[  P_{a},Z_{c}^{\left(  2\right)  }\right]  =0,\\
\left[  Z_{a}^{\left(  1\right)  },Z_{c}^{\left(  1\right)  }\right]   &
=\text{\ }\left[  Z_{a}^{\left(  1\right)  },Z_{c}^{\left(  2\right)
}\right]  =\text{\ }\left[  Z_{a}^{\left(  2\right)  },Z_{c}^{\left(
2\right)  }\right]  =0.
\end{align}

Consider the construction of a three-dimensional Chern-Simons Lagrangian
invariant under $\mathcal{M}_{7}$. \ In fact, Using Theorem $VII.2$ of Ref.
\cite{salg2}, it is possible to show that the only non-vanishing components of
a invariant tensor for the $\mathcal{M}_{7}$ algebra are given by%

\begin{equation}
\left\langle J_{ab}J_{cd}\right\rangle _{\mathcal{M}_{7}}=\alpha_{0}\left(
\eta_{ad}\eta_{bc}-\eta_{ac}\eta_{bd}\right)  ,
\end{equation}%
\begin{equation}
\left\langle J_{ab}Z_{cd}^{\left(  1\right)  }\right\rangle _{\mathcal{M}_{7}%
}=\alpha_{2}\left(  \eta_{ad}\eta_{bc}-\eta_{ac}\eta_{bd}\right)  ,
\end{equation}%
\begin{equation}
\left\langle Z_{ab}^{\left(  1\right)  }Z_{cd}^{\left(  1\right)
}\right\rangle _{\mathcal{M}_{7}}=\left\langle J_{ab}Z_{cd}^{\left(  2\right)
}\right\rangle _{\mathcal{M}_{7}}=\alpha_{4}\left(  \eta_{ad}\eta_{bc}%
-\eta_{ac}\eta_{bd}\right)  ,
\end{equation}%
\begin{equation}
\left\langle P_{a}P_{c}\right\rangle _{\mathcal{M}_{7}}=\alpha_{2}\eta_{ac},
\end{equation}%
\begin{equation}
\left\langle P_{a}Z_{c}^{\left(  1\right)  }\right\rangle _{\mathcal{M}_{7}%
}=\alpha_{4}\eta_{ac},
\end{equation}%
\begin{equation}
\left\langle J_{ab}P_{c}\right\rangle _{\mathcal{M}_{7}}=\alpha_{1}%
\epsilon_{abc},
\end{equation}%
\begin{equation}
\left\langle Z_{ab}^{\left(  1\right)  }P_{c}\right\rangle _{\mathcal{M}_{7}%
}=\left\langle J_{ab}Z_{c}^{\left(  1\right)  }\right\rangle _{\mathcal{M}%
_{7}}=\alpha_{3}\epsilon_{abc},
\end{equation}%
\begin{equation}
\left\langle Z_{ab}^{\left(  1\right)  }Z_{c}^{\left(  1\right)
}\right\rangle _{\mathcal{M}_{7}}=\left\langle J_{ab}Z_{c}^{\left(  2\right)
}\right\rangle _{\mathcal{M}_{7}}=\alpha_{5}\epsilon_{abc}.
\end{equation}
where $\alpha_{0}$, $\alpha_{1}$, $\alpha_{2},\alpha_{3}$, $\alpha_{5}$ and
$\alpha_{5}$ are arbitrary independent constant dimensionless. The $1$-form
gauge connection $A$ $\mathcal{M}_{7}$-valued, is given by,%
\begin{equation}
A=\frac{1}{2}\omega^{ab}J_{ab}+\frac{1}{l}e^{a}P_{a}+\frac{1}{2}k^{\left(
ab,1\right)  }Z_{ab}^{\left(  1\right)  }+\frac{1}{l}h^{\left(  a,1\right)
}Z_{a}^{\left(  1\right)  }+\frac{1}{2}k^{\left(  ab,2\right)  }%
Z_{ab}^{\left(  2\right)  }+\frac{1}{l}h^{\left(  a,2\right)  }Z_{a}^{\left(
2\right)  },
\end{equation}
and the $2$-form curvature is%
\begin{align*}
F  &  =\frac{1}{2}R^{ab}J_{ab}+\frac{1}{l}T^{a}P_{a}+\frac{1}{2}\left(
D_{\omega}k^{\left(  ab,1\right)  }+\frac{1}{l^{2}}e^{a}e^{b}\right)
Z_{ab}^{\left(  1\right)  }+\frac{1}{l}\left(  D_{\omega}h^{\left(
a,1\right)  }+k_{\text{ \ }b}^{a\text{ }\left(  1\right)  }e^{b}\right)
Z_{a}^{\left(  1\right)  }\\
&  +\frac{1}{2}\left(  D_{\omega}k^{\left(  ab,2\right)  }+k_{\text{ \ }%
c}^{a\text{ }\left(  1\right)  }k^{cb\left(  1\right)  }+\frac{1}{l^{2}%
}\left[  e^{a}h^{\left(  b,1\right)  }+h^{\left(  a,1\right)  }e^{b}\right]
\right)  Z_{ab}^{\left(  2\right)  }\\
&  +\frac{1}{l}\left(  D_{\omega}h^{\left(  a,2\right)  }+k_{\text{ \ }%
c}^{a\text{ }\left(  2\right)  \text{ }}e^{c}+k_{\text{ \ }c}^{a\text{
}\left(  1\right)  \text{ }}h^{\left(  c,1\right)  }\right)  Z_{a}^{\left(
2\right)  }%
\end{align*}

Using the dual procedure of S-expansion, we find that the $3$-dimensional
Chern-Simons Lagrangian invariant under the $\mathcal{M}_{7}$-algebra is given
by%
\begin{align}
L_{\text{ }CS\text{ }\left(  2+1\right)  }^{\mathcal{M}_{7}}  &
=\frac{\mathbf{\alpha}_{1}}{l}\varepsilon_{abc}\left(  R^{ab}e^{c}-d\left(
\frac{1}{2}\omega^{ab}e^{c}\right)  \right) \nonumber\\
&  +\frac{\mathbf{\alpha}_{3}}{l}\varepsilon_{abc}\left(  R^{ab}h^{\left(
c,1\right)  }+\mathfrak{R}^{\left(  ab,1\right)  }e^{c}+\frac{1}{3l^{2}}%
e^{a}e^{b}e^{c}-\frac{d}{2}\left(  \omega^{ab}h^{\left(  c,1\right)
}+k^{\left(  ab,1\right)  }e^{c}\right)  \right) \nonumber\\
&  +\frac{\mathbf{\alpha}_{5}}{l}\varepsilon_{abc}\left(  R^{ab}h^{\left(
c,2\right)  }+\mathfrak{R}^{\left(  ab,1\right)  }h^{\left(  c,1\right)
}+\mathfrak{R}^{\left(  ab,2\right)  }e^{c}+\frac{1}{l^{2}}e^{a}%
e^{b}h^{\left(  c,1\right)  }\right. \nonumber\\
&  \left.  -\frac{d}{2}\left(  \omega^{ab}h^{\left(  c,2\right)  }+k^{\left(
ab,1\right)  }h^{\left(  c,1\right)  }+k^{\left(  ab,2\right)  }e^{c}\right)
\right)  +\frac{\mathbf{\alpha}_{0}}{2}\left(  \omega_{\;b}^{a}d\omega
_{\;a}^{b}+\frac{2}{3}\omega_{\hspace{0.05cm}\text{ }b}^{a}\omega
_{\hspace{0.05cm}\text{ }c}^{b}\omega_{\hspace{0.05cm}\text{ }a}^{c}\right)
\nonumber\\
&  +\frac{\mathbf{\alpha}_{2}}{2}\left(  \omega_{\;b}^{a}dk_{\;a}^{b\text{
}\left(  1\right)  }+k_{\;b}^{a\text{ }\left(  1\right)  }d\omega_{\;a}%
^{b}+2\omega_{\hspace{0.05cm}\text{ }b}^{a}\omega_{\hspace{0.05cm}\text{ }%
c}^{b}k_{\hspace{0.05cm}\text{ }a}^{c\text{ }\left(  1\right)  }+\frac
{2}{l^{2}}e_{a}T^{a}\right) \nonumber\\
&  +\frac{\mathbf{\alpha}_{4}}{2}\left(  \omega_{\;b}^{a}dk_{\;a}^{b\text{
}\left(  2\right)  }+k_{\;b}^{a~\left(  2\right)  }d\omega_{\;a}^{b}%
+2\omega_{\hspace{0.05cm}\text{ }b}^{a}\omega_{\hspace{0.05cm}\text{ }c}%
^{b}k_{\hspace{0.05cm}\text{ }a}^{c\text{ }\left(  2\right)  }+k_{\;b}%
^{a\text{ }\left(  1\right)  }dk_{\;a}^{b\text{ }\left(  1\right)  }%
+2\omega_{\hspace{0.05cm}\text{ }b}^{a}k_{\hspace{0.05cm}\text{ }c}^{b~\left(
1\right)  }k_{\hspace{0.05cm}\text{ }a}^{c\text{ }\left(  1\right)  }\right.
\nonumber\\
&  \left.  +\frac{2}{l^{2}}e_{a}\mathfrak{T}^{\left(  a,1\right)  }+\frac
{2}{l^{2}}h_{a}^{\text{ }\left(  1\right)  }T^{a}\right)  . \label{LBCS3}%
\end{align}
where%
\begin{align}
\mathfrak{R}^{\left(  ab,1\right)  }  &  =D_{\omega}k^{\left(  ab,1\right)
},\\
\mathfrak{R}^{\left(  ab,2\right)  }  &  =D_{\omega}k^{\left(  ab,2\right)
}+k_{\text{ \ }c}^{a\text{ }\left(  1\right)  }k^{cb\left(  1\right)  }\\
\mathfrak{T}^{\left(  a,1\right)  }  &  =D_{\omega}h^{\left(  a,1\right)
}+k_{\text{ \ }c}^{a\text{ }\left(  1\right)  \text{ }}e^{c},
\end{align}

The Lagrangian $\left(  \text{\ref{LBCS3}}\right)  $ is split into six
independent pieces, each one proportional to $\alpha_{1}$, $\alpha_{3}$,
$\alpha_{5}$, $\alpha_{0}$, $\alpha_{2}$, $\alpha_{4}$. \ The term
proportional to $\alpha_{1}$ corresponds to the Chern-Simons Lagrangian for
$ISO\left(  2,1\right)  $ which contains the Eintein-Hilbert term
$\varepsilon_{abc}R^{ab}e^{c}$. \ 

\ Varying the Lagrangian $\left(  \text{\ref{LBCS3}}\right)  $ we have%
\begin{align*}
\delta L_{\text{ }CS\text{ }\left(  2+1\right)  }^{\mathcal{M}_{7}}  &
=\frac{1}{l}\varepsilon_{abc}\left(  \alpha_{1}R^{ab}+\frac{\alpha_{3}}{l^{2}%
}e^{a}e^{b}+\alpha_{3}\mathfrak{R}^{\left(  ab,1\right)  }+\mathfrak{R}%
^{\left(  ab,2\right)  }\right)  \delta e^{c}\\
&  +\frac{1}{l}\varepsilon_{abc}\left(  \alpha_{3}R^{ab}+\alpha_{5}%
\mathfrak{R}^{\left(  ab,1\right)  }+\frac{\alpha_{5}}{l^{2}}e^{a}%
e^{b}\right)  \delta h^{\left(  c,1\right)  }\\
&  +\frac{1}{l}\varepsilon_{abc}\left(  \alpha_{5}R^{ab}\right)  \delta
h^{\left(  c,2\right)  }+\frac{1}{l}\varepsilon_{abc}\delta\omega^{ab}\left(
\alpha_{1}T^{c}+\alpha_{3}D_{\omega}h^{\left(  c,1\right)  }+\alpha
_{5}D_{\omega}h^{\left(  c,2\right)  }\right) \\
&  +\frac{1}{l}\varepsilon_{acd}\delta\omega^{ab}\left(  \alpha_{3}%
e_{b}k^{\left(  cd,1\right)  }+\alpha_{5}h_{b}^{\text{ },\left(  1\right)
}k^{\left(  cd,1\right)  }+\alpha_{5}e_{b}k^{\left(  cd,2\right)  }\right) \\
&  +\frac{1}{l}\varepsilon_{abc}\delta k^{\left(  ab,1\right)  }\left(
\alpha_{3}T^{c}+\alpha_{5}D_{\omega}h^{\left(  c,1\right)  }\right)  +\frac
{1}{l}\varepsilon_{acd}\delta k^{\left(  ab,1\right)  }\left(  2\alpha
_{5}k_{b}^{\text{ }c,\left(  1\right)  }e^{d}\right) \\
&  +\frac{1}{l}\varepsilon_{abc}\delta k^{\left(  ab,2\right)  }\left(
\alpha_{5}T^{c}\right)  +\frac{\alpha_{0}}{2}\left(  \delta L_{3}%
^{Lorentz}\right)  +\frac{\alpha_{2}}{2}\left(  \delta L_{3}^{Lorentz}\left(
k^{\left(  1\right)  }\right)  \right) \\
&  +\frac{\alpha_{4}}{2}\left(  \delta L_{3}^{Lorentz}\left(  k^{\left(
2\right)  }\right)  \right)  +\frac{\alpha_{4}}{2}\left(  \delta
L_{3}^{Lorentz}\left(  k^{\left(  1\right)  }k^{\left(  1\right)  }\right)
\right) \\
&  +\delta e_{a}\left(  \frac{\alpha_{4}}{l^{2}}\mathfrak{T}^{\left(
a,1\right)  }+\frac{2\alpha_{2}}{l^{2}}T^{a}\right)  +\delta\omega^{ab}\left(
\frac{\alpha_{2}}{l^{2}}e_{a}e_{b}+\frac{\alpha_{4}}{l^{2}}e_{b}h_{a}^{\text{
}\left(  1\right)  }\right) \\
&  +\delta h_{a}^{\text{ },\left(  1\right)  }\left(  \frac{2\alpha_{4}}%
{l^{2}}T^{a}\right)  +\delta k^{\left(  ab,1\right)  \text{ }}\left(
\frac{\alpha_{4}}{l^{2}}e_{b}e_{a}\right)  .
\end{align*}
where $L_{3}^{Lorentz}=\omega d\omega+\frac{2}{3}\omega^{3}.$

If we consider the case where $k^{\left(  ab,1\right)  }=$ $k^{\left(
ab,2\right)  }=0$, $h^{\left(  a,1\right)  }=0$ and $h^{\left(  a,2\right)
}=0)$ with the condition $\alpha_{1}=\alpha_{3}=\alpha_{5}=0$ we have%
\begin{align*}
\delta L_{\text{ }CS\text{ }\left(  2+1\right)  }^{\mathcal{M}_{7}}  &
=\frac{\alpha_{0}}{2}\left(  \delta L_{3}^{Lorentz}\right)  +\frac{\alpha_{2}%
}{2l^{2}}\delta\omega^{ab}\left(  e_{a}e_{b}\right)  +\frac{\alpha_{2}}%
{2l^{2}}\delta e^{a}\left(  T_{a}\right) \\
&  =\mathbf{\alpha}_{0}\delta\omega^{ab}\left(  R_{ab}\right)  +\frac
{\alpha_{2}}{2l^{2}}\delta\omega^{ab}\left(  e_{a}e_{b}\right)  +\frac
{\alpha_{2}}{2l^{2}}\delta e^{a}\left(  T_{a}\right)  .
\end{align*}
Choosing $\alpha_{0}=$ $\alpha_{2}$ we have that $\delta L_{\text{ }CS\text{
}\left(  2+1\right)  }^{\mathcal{M}_{7}}=0$, leads to the following equations
of motion%

\begin{align}
R^{ab}+\frac{1}{l^{2}}e^{a}e^{b}  &  =0,\\
T_{a}  &  =0.
\end{align}
which correspond to the equations of general relativity with cosmological
constant in $\left(  2+1\right)  $-dimensions.

\subsection{$(4+1)$\textbf{-dimensional Chern-Simons Lagrangian invariant
under }$\mathcal{M}_{7}$-\textbf{algebra}}

The only non-vanishing components of a invariant tensor for the $\mathcal{M}%
_{7}$ algebra are given by

In $D=5$, the only non-vanishing components of a invariant tensor for the
$\mathcal{M}_{7}$ algebra are given by%
\begin{equation}
\left\langle J_{ab}J_{cd}P_{f}\right\rangle _{\mathcal{M}_{7}}=\frac{4}%
{3}l^{3}\alpha_{1}\epsilon_{abcdf}.
\end{equation}%
\begin{equation}
\left\langle J_{ab}J_{cd}Z_{f}^{\left(  1\right)  }\right\rangle
_{\mathcal{M}_{7}}=\frac{4}{3}l^{3}\alpha_{3}\epsilon_{abcdf}.
\end{equation}%
\begin{equation}
\left\langle J_{ab}Z_{cd}^{\left(  1\right)  }P_{f}\right\rangle
_{\mathcal{M}_{7}}=\frac{4}{3}l^{3}\alpha_{3}\epsilon_{abcdf}.
\end{equation}%
\begin{equation}
\left\langle J_{ab}J_{cd}Z_{f}^{\left(  2\right)  }\right\rangle
_{\mathcal{M}_{7}}=\frac{4}{3}l^{3}\alpha_{5}\epsilon_{abcdf}.
\end{equation}%
\begin{equation}
\left\langle J_{ab}Z_{cd}^{\left(  1\right)  }Z_{f}^{\left(  1\right)
}\right\rangle _{\mathcal{M}_{7}}=\frac{4}{3}l^{3}\alpha_{5}\epsilon_{abcdf}.
\end{equation}
where $\alpha_{1}$, $\alpha_{3}$ and $\alpha_{5}$ are arbitrary independent
constant of dimensions $\left[  length\right]  ^{-3}$. Using the dual
procedure of S-expansion, we find that the $5$-dimensional Chern-Simons
Lagrangian invariant under the $\mathcal{M}_{7}$-algebra is given by%

\begin{align}
L_{\text{ }\left(  4+1\right)  }^{\mathcal{M}_{7}}  &  =\alpha_{1}%
\varepsilon_{abcdf}\left(  l^{2}R^{ab}R^{cd}e^{f}\right) \nonumber\\
&  \alpha_{3}\varepsilon_{abcdf}\left(  l^{2}R^{ab}R^{cd}h^{\left(
f,1\right)  }+2l^{2}R^{ab}\mathfrak{R}^{\left(  cd,1\right)  }e^{f}+\frac
{2}{3}R^{ab}e^{c}e^{d}e^{f}\right) \nonumber\\
&  \alpha_{5}\varepsilon_{abcdf}\left(  l^{2}R^{ab}R^{cd}h^{\left(
f,2\right)  }+2l^{2}R^{ab}\mathfrak{R}^{\left(  cd,1\right)  }h^{\left(
f,1\right)  }+2l^{2}R^{ab}\mathfrak{R}^{\left(  cd,2\right)  }e^{f}\right.
\nonumber\\
&  \left.  +l^{2}\mathfrak{R}^{\left(  ab,1\right)  }\mathfrak{R}^{\left(
cd,1\right)  }e^{f}+2R^{ab}e^{c}e^{d}h^{\left(  f,1\right)  }+\frac{2}%
{3}\mathfrak{R}^{\left(  ab,1\right)  }e^{c}e^{d}e^{f}+\frac{1}{5l^{2}}%
e^{a}e^{b}e^{c}e^{d}e^{f}\right)  \label{cuatr}%
\end{align}
Varying the Lagrangian $\left(  \text{\ref{cuatr}}\right)  $ we have%

\begin{align*}
\delta L_{\text{ }\left(  4+1\right)  }^{\mathcal{M}_{7}}  &  =\varepsilon
_{abcdf}\left(  \mathbf{\alpha}_{1}l^{2}R^{ab}R^{cd}+2\mathbf{\alpha}_{3}%
l^{2}R^{ab}\mathfrak{R}^{\left(  cd,1\right)  }+2\mathbf{\alpha}_{3}%
R^{ab}e^{c}e^{d}+2\mathbf{\alpha}_{5}l^{2}R^{ab}\mathfrak{R}^{\left(
cd,2\right)  }\right. \\
&  \left.  +\mathbf{\alpha}_{5}l^{2}\mathfrak{R}^{\left(  ab,1\right)
}\mathfrak{R}^{\left(  cd,1\right)  }+4\mathbf{\alpha}_{5}R^{ab}%
e^{c}h^{\left(  d,1\right)  }+2\mathbf{\alpha}_{5}\mathfrak{R}^{\left(
ab,1\right)  }e^{c}e^{d}+\frac{1}{l^{2}}\mathbf{\alpha}_{5}e^{a}e^{b}%
e^{c}e^{d}\right)  \delta e^{f}\\
&  +\varepsilon_{abcdf}\left(  \mathbf{\alpha}_{3}l^{2}R^{ab}R^{cd}%
+2\mathbf{\alpha}_{5}l^{2}R^{ab}\mathfrak{R}^{\left(  cd,1\right)
}+2\mathbf{\alpha}_{5}R^{ab}e^{c}e^{d}\right)  \delta h^{\left(  f,1\right)
}\\
&  +\varepsilon_{abcdf}\mathbf{\alpha}_{5}l^{2}R^{ab}R^{cd}\delta h^{\left(
f,2\right)  }+2\varepsilon_{abcdf}\mathbf{\alpha}_{5}l^{2}\delta k^{\left(
ab,2\right)  }R^{cd}T^{f}\\
&  +\varepsilon_{abcdf}\delta k^{\left(  ab,1\right)  }(2\mathbf{\alpha}%
_{3}l^{2}R^{cd}T^{f}+2\mathbf{\alpha}_{5}l^{2}R^{cd}D_{\omega}h^{\left(
f,1\right)  }+2\mathbf{\alpha}_{5}e^{c}e^{d}T^{f}\\
&  +2\mathbf{\alpha}_{5}l^{2}D_{\omega}k^{\left(  cd,1\right)  }%
T^{f})+\varepsilon_{acdfg}\delta k^{\left(  ab,1\right)  }\left(  4\alpha
_{5}l^{2}k_{b}^{\text{ }c,\left(  1\right)  }R^{df}e^{g}+2\alpha_{5}l^{2}%
R_{b}^{\text{ }c}k^{\left(  df,1\right)  }T^{g}\right) \\
&  +\varepsilon_{abcdf}\delta\omega^{ab}\left[  2\mathbf{\alpha}_{1}%
l^{2}R^{cd}T^{f}+2\mathbf{\alpha}_{3}l^{2}R^{cd}D_{\omega}h^{\left(
f,1\right)  }+2\mathbf{\alpha}_{3}l^{2}\mathfrak{R}^{\left(  cd,1\right)
}T^{f}\right. \\
&  \left.  -2\alpha_{3}l^{2}k^{\left(  cd,1\right)  }R^{fg}e_{g}%
+2\mathbf{\alpha}_{3}e^{c}e^{d}T^{f}+2\mathbf{\alpha}_{5}l^{2}R^{cd}D_{\omega
}h^{\left(  f,2\right)  }+2\mathbf{\alpha}_{5}l^{2}\mathfrak{R}^{\left(
cd,1\right)  }D_{\omega}h^{\left(  f,1\right)  }\right. \\
&  \left.  -2\alpha_{5}l^{2}k^{\left(  cd,1\right)  }R^{fg}h_{g}^{\text{
},\left(  1\right)  }+2\mathbf{\alpha}_{5}l^{2}D_{\omega}k^{\left(
cd,2\right)  }T^{f}-2\alpha_{5}l^{2}k^{\left(  cd,2\right)  }R^{fg}%
e_{g}\right. \\
&  \left.  +4\alpha_{5}l^{2}\mathfrak{R}_{\text{ }g}^{c\text{ },\left(
1\right)  }k^{\left(  gd,1\right)  }e^{f}+2\alpha_{5}l^{2}k_{\text{ }%
g}^{c\text{ },\left(  1\right)  }k^{\left(  gd,1\right)  }T^{f}%
+4\mathbf{\alpha}_{5}e^{c}T^{d}h^{\left(  f,1\right)  }+2\mathbf{\alpha}%
_{5}e^{c}e^{d}D_{\omega}h^{\left(  f,1\right)  }\right] \\
&  +\varepsilon_{acdfg}\delta\omega^{ab}\left(  2\alpha_{3}l^{2}e_{b}%
R^{cd}k^{\left(  fg,1\right)  }+2\alpha_{5}l^{2}h_{b}^{\text{ },\left(
1\right)  }R^{cd}k^{\left(  fg,1\right)  }+2\alpha_{5}l^{2}e_{b}%
R^{cd}k^{\left(  fg,2\right)  }\right. \\
&  \left.  -2\alpha_{5}l^{2}\mathfrak{R}_{b}^{\text{ }c,\left(  1\right)
}k^{\left(  df,1\right)  }e^{g}+2\alpha_{5}k_{b}^{\text{ }c,\left(  1\right)
}\mathfrak{R}^{\left(  df,1\right)  }e^{g}+e_{b}k^{\left(  cd,1\right)
}\mathfrak{R}^{\left(  fg,1\right)  }+2\alpha_{5}e_{b}k^{\left(  cd,1\right)
}e^{f}e^{g}\right)
\end{align*}

When a solution without matter $\left(  k^{\left(  ab,1\right)  }=0,k^{\left(
ab,2\right)  }=0,h^{\left(  a,1\right)  }=0,h^{\left(  a,2\right)  }=0\right)
$ is singled out, we are left with%
\begin{align*}
\delta L_{\text{ }\left(  4+1\right)  }^{\mathcal{M}_{7}} &  =\varepsilon
_{abcdf}\left[  \left(  \mathbf{\alpha}_{1}l^{2}R^{ab}R^{cd}+2\mathbf{\alpha
}_{3}R^{ab}e^{c}e^{d}+\frac{1}{l^{2}}\mathbf{\alpha}_{5}e^{a}e^{b}e^{c}%
e^{d}\right)  \delta e^{f}\right.  \\
&  \left.  +\left(  \mathbf{\alpha}_{3}l^{2}R^{ab}R^{cd}+2\mathbf{\alpha}%
_{5}R^{ab}e^{c}e^{d}\right)  \delta h^{\left(  f,1\right)  }+2\mathbf{\alpha
}_{5}l^{2}\delta k^{\left(  ab,2\right)  }R^{cd}T^{f}+\mathbf{\alpha}_{5}%
l^{2}R^{ab}R^{cd}\delta h^{\left(  f,2\right)  }\right.  \\
&  \left.  +\delta k^{\left(  ab,1\right)  }\left(  2\mathbf{\alpha}_{3}%
l^{2}R^{cd}T^{f}+2\mathbf{\alpha}_{5}e^{c}e^{d}T^{f}\right)  +\delta
\omega^{ab}\left(  2\mathbf{\alpha}_{1}l^{2}R^{cd}T^{f}+2\mathbf{\alpha}%
_{3}e^{c}e^{d}T^{f}\right)  \right]
\end{align*}
So when $\alpha_{1}$ and $\alpha_{5}$ vanish we finally get
\begin{align}
\delta L_{\text{ }\left(  4+1\right)  }^{\mathcal{M}_{7}} &  =\varepsilon
_{abcdf}\left(  2\mathbf{\alpha}_{3}R^{ab}e^{c}e^{d}\right)  \delta
e^{f}+\varepsilon_{abcdf}\left(  \mathbf{\alpha}_{3}l^{2}R^{ab}R^{cd}\right)
\delta h^{\left(  f,1\right)  }\nonumber\\
&  +\varepsilon_{abcdf}\delta k^{\left(  ab,1\right)  }\left(  2\mathbf{\alpha
}_{3}l^{2}R^{cd}T^{f}\right)  +\delta\omega^{ab}\left(  2\mathbf{\alpha}%
_{3}e^{c}e^{d}T^{f}\right)  \label{evelyn1}%
\end{align}

Therefore, if we impose the torsionless condition, we see that the
Chern-Simons Lagrangian in $D=5$ invariant under $M_{7}$ leads to the same
equations of motion than the Chern-Simons Lagrangian in $D=5$ invariant under
$M_{5}$ \cite{salg1}. From (\ref{evelyn1}), like in Ref. \cite{salg1}, we can
see that in the limit where $l=0$ the extra constraints just vanish, and
$\delta L_{CS}=0$ leads us to the Einstein-Hilbert dynamics in vacuum,%
\begin{equation}
\delta L_{CS\;\left(  4+1\right)  }^{\mathcal{M}_{7}}=\varepsilon
_{abcdf}\left(  2\mathbf{\alpha}_{3}R^{ab}e^{c}e^{d}\right)  \delta
e^{f}+\varepsilon_{abcdf}\delta\omega^{ab}\left(  2\mathbf{\alpha}_{3}%
e^{c}e^{d}T^{f}\right)  .
\end{equation}
Similarly, when the cosmological constant is not considered and a solution
without matter is singled out, the strict limit where the coupling constant
$l$ equals zero yields just to the Einstein Hilbert term in the Lagrangian%
\begin{equation}
L_{CS\;\left(  4+1\right)  }^{\mathcal{M}_{7}}=\frac{2}{3}\alpha
_{3}\varepsilon_{abcdf}R^{ab}e^{c}e^{d}e^{f}.
\end{equation}

\subsection{$(6+1)$\textbf{-dimensional Chern-Simons Lagrangian invariant
under }$\mathcal{M}_{5}$-\textbf{algebra}}

\ Now, consider a Chern-Simons action $(6+1)$-dimensional invariant under
$\mathcal{M}_{5}$-algebra.The $1$-form gauge connection $A$ $\mathcal{M}_{5}%
$-valued, is given by%

\begin{equation}
A=\frac{1}{2}\omega^{ab}J_{ab}+\frac{1}{l}e^{a}P_{a}+\frac{1}{2}k^{ab}%
Z_{ab}+\frac{1}{l}h^{a}Z_{a},
\end{equation}
and the $2$-forma curvature is given by%
\begin{equation}
F=\frac{1}{2}R^{ab}J_{ab}+\frac{1}{l}T^{a}P_{a}+\frac{1}{2}\left(  D_{\omega
}k^{ab}+\frac{1}{l^{2}}e^{a}e^{b}\right)  Z_{ab}+\frac{1}{l}\left(  D_{\omega
}h^{a}+k_{\text{ \ }b}^{a}e^{b}\right)  Z_{a}.
\end{equation}

Using the dual procedure of S-expansion, we find that the $7$-dimensional
Chern-Simons Lagrangian invariant under the $\mathcal{M}_{5}$-algebra is given
by
\begin{align}
L_{\;\left(  6+1\right)  }^{\mathcal{M}_{5}}  &  =\frac{\alpha_{1}}%
{l}\varepsilon_{abcdefg}\left(  R^{ab}R^{cd}R^{ef}e^{g}\right) \nonumber\\
&  +\frac{\alpha_{3}}{l}\varepsilon_{abcdefg}\left(  R^{ab}R^{cd}R^{ef}%
h^{g}+3R^{ab}R^{cd}D_{\omega}k^{ef}e^{g}+\frac{1}{l^{2}}R^{ab}R^{cd}e^{e}%
e^{f}e^{g}\right)  .
\end{align}
where $\alpha_{1}=\lambda_{1}\kappa,$ $\alpha_{3}=\lambda_{3}\kappa.$ \ From
here we see that the Einstein-Hilbert term is not present in the Lagrangian.
This result holds for all $D=p$-dimensional Chern-Simons Lagrangian invariant
under an algebra $\mathcal{M}_{m}$ if $p>m$.

Varying the Lagrangian we have%

\begin{align}
\delta L_{\;\left(  6+1\right)  }^{\mathcal{M}_{5}}  &  =\frac{\mathbf{1}}%
{l}\varepsilon_{abcdefg}\left(  \alpha_{1}R^{ab}R^{cd}R^{ef}+3\alpha_{3}%
R^{ab}R^{cd}D_{\omega}k^{ef}+\frac{3}{l^{2}}\alpha_{3}R^{ab}R^{cd}e^{e}%
e^{f}\right)  \delta e^{g}\nonumber\\
&  +\frac{\mathbf{1}}{l}\varepsilon_{abcdefg}\left(  \alpha_{3}R^{ab}%
R^{cd}R^{ef}\right)  \delta h^{g}\nonumber\\
&  +\frac{\mathbf{1}}{l}\varepsilon_{abcdefg}\delta\omega^{ab}\left(
3\alpha_{1}R^{cd}R^{ef}T^{g}+3\alpha_{3}R^{cd}R^{ef}\mathfrak{\ }D_{\omega
}h^{g}+6\alpha_{3}R^{cd}D_{\omega}k^{ef}T^{g}\right. \nonumber\\
&  \left.  +\frac{6}{l^{2}}\alpha_{3}R^{cd}e^{e}e^{f}T^{g}\right)
+\frac{\mathbf{1}}{l}\varepsilon_{acdefgh}\delta\omega^{ab}\left(
3e_{b}R^{cd}R^{ef}k^{gh}\right) \nonumber\\
&  +\frac{\mathbf{1}}{l}\varepsilon_{abcdefg}\delta k^{ab}\left(  3\alpha
_{3}R^{cd}R^{ef}T^{g}\right)  .
\end{align}
from which we can see it is not possible to obtain the Einstein-Hilbert dynamics.

\textbf{In fact}, imposing the torsionless condition and if we consider the
case where $k^{ab}=0$ , $h^{a}=0$ with $\alpha_{1}=0$ we find%

\begin{equation}
\delta L_{\;\left(  6+1\right)  }^{\mathfrak{B}_{5}}=\frac{\alpha_{3}}{l^{2}%
}\varepsilon_{abcdefg}R^{ab}R^{cd}e^{e}e^{f}\delta e^{g}+\frac{\alpha_{3}}%
{l}\varepsilon_{abcdefg}R^{ab}R^{cd}R^{ef}\delta h^{g},
\end{equation}
which obviously does not correspond to the dynamics of General Relativity.

\subsection{$(6+1)$\textbf{-dimensional Chern-Simons Lagrangian invariant
under }$\mathcal{M}_{7}$-\textbf{algebra}}

Consider the $7$-dimensional Chern-Simons Lagrangian invariant under the $AdS$
algebra
\begin{align*}
L_{CS\;\left(  6+1\right)  }^{AdS}  &  =\kappa\left[  \varepsilon
_{abcdefg}\left(  \frac{1}{l}R^{ab}R^{cd}R^{ef}e^{g}+\frac{1}{l^{3}}%
R^{ab}R^{cd}e^{e}e^{f}e^{g}+\frac{3}{5l^{5}}R^{ab}e^{c}e^{d}e^{e}e^{f}%
e^{g}+\frac{1}{7l^{7}}e^{a}e^{b}e^{c}e^{d}e^{e}e^{f}e^{g}\right)  \right] \\
&  +\beta_{2,2}\left[  R_{\;b}^{a}R_{\;a}^{b}+\frac{2}{l^{2}}\left(
T^{a}T_{a}-R^{ab}e_{a}e_{b}\right)  \right]  \left(  \omega_{\;d}^{c}%
d\omega_{\;c}^{d}+\frac{2}{3}\omega_{\;f\;}^{c}\omega_{\;g}^{f}\omega
_{\;c}^{g}+\frac{2}{l^{2}}e_{c}T^{c}\right) \\
&  +\beta_{4}\left[  \left(  \omega_{\;b}^{a}d\omega_{\;c}^{b}d\omega
_{\;d}^{c}d\omega_{\;a}^{d}+\frac{8}{5}\omega_{\;b}^{a}\omega_{\;c}^{b}%
\omega_{\;d}^{c}d\omega_{\;e}^{d}d\omega_{\;a}^{e}+\frac{4}{5}\omega_{\;b}%
^{a}d\omega_{\;c}^{b}\omega_{\;d}^{c}\omega_{\;e}^{d}d\omega_{\;a}^{e}\right.
\right. \\
&  \left.  +2\omega_{\;b}^{a}\omega_{\;c}^{b}\omega_{\;d}^{c}\omega_{\;e}%
^{d}\omega_{\;f}^{e}d\omega_{\;a}^{f}+\frac{4}{7}\omega_{\;b}^{a}\omega
_{\;c}^{b}\omega_{\;d}^{c}\omega_{\;e}^{d}\omega_{\;f}^{e}\omega_{\;g}%
^{f}\omega_{\;a}^{g}\right) \\
&  \left.  +\frac{1}{l^{2}}4T_{a}R_{\;b}^{a}R_{\;c}^{b}e^{c}+\frac{1}{l^{4}%
}\left[  2\left(  R^{ab}e_{a}e_{b}+T^{a}T_{a}\right)  T^{c}e_{c}\right]
\right]  .
\end{align*}

Using the dual procedure of S-expansion, we find that the $7$-dimensional
Chern-Simons Lagrangian invariant under the $\mathcal{M}_{7}$-algebra is given by%

\begin{align}
&  L_{CS\;\left(  6+1\right)  }^{\mathcal{M}_{7}}\nonumber\\
&  =\mathbf{\alpha}_{1}l^{4}\varepsilon_{abcdefg}R^{ab}R^{cd}R^{ef}%
e^{g}+\mathbf{\alpha}_{3}\varepsilon_{abcdefg}\left(  l^{4}R^{ab}R^{cd}%
R^{ef}h^{\left(  g,1\right)  }+3l^{4}R^{ab}R^{cd}\mathfrak{R}^{\left(
ef,1\right)  }e^{g}+l^{2}R^{ab}R^{cd}e^{e}e^{f}e^{g}\right) \nonumber\\
&  +\mathbf{\alpha}_{5}\varepsilon_{abcdefg}\left(  l^{4}R^{ab}R^{cd}%
R^{ef}h^{\left(  g,2\right)  }+3l^{4}R^{ab}\mathfrak{R}^{\left(  cd,1\right)
}\mathfrak{R}^{\left(  ef,1\right)  }e^{g}+3l^{4}R^{ab}R^{cd}\mathfrak{R}%
^{\left(  ef,2\right)  }e^{g}\right. \nonumber\\
&  \left.  +3l^{4}R^{ab}R^{cd}\mathfrak{R}^{\left(  ef,1\right)  }h^{\left(
g,1\right)  }+2l^{2}R^{ab}\mathfrak{R}^{\left(  cd,1\right)  }e^{e}e^{f}%
e^{g}+3l^{2}R^{ab}R^{cd}e^{e}e^{f}h^{\left(  g,1\right)  }+\frac{3}{5}%
R^{ab}e^{c}e^{d}e^{e}e^{f}e^{g}\right) \nonumber\\
&  +\mathbf{\alpha}_{0\left\{  2,2\right\}  }l^{5}\left[  \left(  R_{\;b}%
^{a}R_{\;a}^{b}\right)  L_{3}^{Lorentz}\right] \nonumber\\
&  +\mathbf{\alpha}_{2\left\{  2,2\right\}  }l^{5}\left[  \left(  R_{\;b}%
^{a}R_{\;a}^{b}\right)  \left(  L_{3}^{Lorentz}\left(  k^{\left(  1\right)
}\right)  +\frac{2}{l^{2}}e_{c}T^{c}\right)  +2\left(  R_{\;b}^{a}%
\mathfrak{R}_{\;a}^{b\text{ }\left(  1\right)  }\right)  L_{3}^{Lorentz}%
\right. \nonumber\\
&  \left.  +\frac{2}{l^{2}}\left(  T^{a}T_{a}-R^{ab}e_{a}e_{b}\right)
L_{3}^{Lorentz}\right] \nonumber\\
&  +\mathbf{\alpha}_{4\left\{  2,2\right\}  }l^{5}\left[  \left(  R_{\;b}%
^{a}R_{\;a}^{b}\right)  \left(  L_{3}^{Lorentz}\left(  k^{\left(  2\right)
}\right)  +L_{3}^{Lorentz}\left(  k^{\left(  1\right)  }k^{\left(  1\right)
}\right)  +\frac{2}{l^{2}}e_{c}\mathfrak{T}^{c\text{ }\left(  1\right)
}+\frac{2}{l^{2}}h_{c}^{\text{ }\left(  1\right)  }T^{c}\right)  \right.
\nonumber\\
&  \left.  +2\left(  R_{\;b}^{a}\mathfrak{R}_{\;a}^{b\text{ }\left(  1\right)
}\right)  \left(  L_{3}^{Lorentz}\left(  k^{\left(  1\right)  }\right)
+\frac{2}{l^{2}}e_{c}T^{c}\right)  +\left(  \mathfrak{R}_{\;b}^{a\text{
}\left(  1\right)  }\mathfrak{R}_{\;a}^{b\text{ }\left(  1\right)  }\right)
L_{3}^{Lorentz}\right. \nonumber\\
&  \left.  +2\left(  R_{\;b}^{a}\mathfrak{R}_{\;a}^{b\text{ }\left(  2\right)
}\right)  L_{3}^{Lorentz}+\frac{2}{l^{2}}\left(  T^{a}T_{a}-R^{ab}e_{a}%
e_{b}\right)  \left(  L_{3}^{Lorentz}\left(  k^{\left(  1\right)  }\right)
+\frac{2}{l^{2}}e_{c}T^{c}\right)  \right. \nonumber\\
&  \left.  +\frac{2}{l^{2}}\left(  2T^{a}\mathfrak{T}_{a}^{\text{ }\left(
1\right)  }-2R^{ab}e_{a}h_{b}^{\text{ }\left(  1\right)  }-\mathfrak{R}%
^{\left(  ab,1\right)  }e_{a}e_{b}\right)  L_{3}^{Lorentz}\right] \nonumber\\
&  +\mathbf{\alpha}_{0\left\{  4\right\}  }l^{5}\left[  L_{7}^{Lorentz}%
\right]  +\mathbf{\alpha}_{2\left\{  4\right\}  }l^{5}\left[  L_{7}%
^{Lorentz}\left(  k^{\left(  1\right)  }\right)  +\frac{1}{l^{2}}4T_{a}%
R_{\;b}^{a}R_{\;c}^{b}e^{c}\right] \nonumber\\
&  +\mathbf{\alpha}_{4\left\{  4\right\}  }l^{5}\left[  L_{7}^{Lorentz}\left(
k^{\left(  2\right)  }\right)  +L_{7}^{Lorentz}\left(  k^{\left(  1\right)
}k^{\left(  1\right)  }\right)  \right. \nonumber\\
&  \left.  +\frac{4}{l^{2}}\left(  T_{a}R_{\;b}^{a}R_{\;c}^{b}h^{\left(
c,1\right)  }+\mathfrak{T}_{a}^{\text{ }\left(  1\right)  }R_{\;b}^{a}%
R_{\;c}^{b}e^{c}+T_{a}R_{\;b}^{a}\mathfrak{R}_{\;c}^{b\text{ }\left(
1\right)  }e^{c}+T_{a}\mathfrak{R}_{\;b}^{a\text{ }\left(  1\right)  }%
R_{\;c}^{b}e^{c}\right)  \right. \nonumber\\
&  \left.  +\frac{1}{l^{4}}\left[  2\left(  R^{ab}e_{a}e_{b}+T^{a}%
T_{a}\right)  T^{c}e_{c}\right]  \right]  . \label{l7b7}%
\end{align}

The Lagrangian $\left(  \text{\ref{l7b7}}\right)  $ is split into nine
independent pieces, each one proportional to $\alpha_{1}$, $\alpha_{3}$,
$\alpha_{5}$, $\alpha_{0\left\{  2,2\right\}  }$, $\alpha_{2\left\{
2,2\right\}  }$, $\alpha_{4\left\{  2,2\right\}  }$, $\alpha_{0\left\{
4\right\}  }$, $\alpha_{2\left\{  4\right\}  }$ and $\alpha_{4\left\{
4\right\}  }$. \ The term proportional to $\alpha_{1}$ corresponds to the
Chern-Simons Lagrangian for $\mathrm{ISO}\left(  6,1\right)  $ group. The
Eintein-Hilbert term $\varepsilon_{abcdefg}R^{ab}e^{c}e^{d}e^{e}e^{f}e^{g}$
appears in the term proportional to $\alpha_{5}.$ \ 

\ Varying the Lagrangian $\left(  \text{\ref{l7b7}}\right)  $ for the case
$\alpha_{0\left\{  2,2\right\}  }$ $=$ $\alpha_{2\left\{  2,2\right\}  }$ $=$
$\alpha_{4\left\{  2,2\right\}  }$ $=$ $\alpha_{0\left\{  4\right\}  }=$
$\alpha_{2\left\{  4\right\}  }$ $=$ $\alpha_{4\left\{  4\right\}  }$ $=0$, we have%

\begin{align*}
\delta L_{CS\;\left(  6+1\right)  }^{\mathcal{M}_{7}}  &  =\varepsilon
_{abcdefg}\left(  \mathbf{\alpha}_{1}l^{4}R^{ab}R^{cd}R^{ef}+3\mathbf{\alpha
}_{3}l^{4}R^{ab}R^{cd}\mathfrak{R}^{\left(  ef,1\right)  }+3\mathbf{\alpha
}_{3}l^{2}R^{ab}R^{cd}e^{e}e^{f}\right. \\
&  \left.  +3\mathbf{\alpha}_{5}l^{4}R^{ab}\mathfrak{R}^{\left(  cd,1\right)
}\mathfrak{R}^{\left(  ef,1\right)  }+3\mathbf{\alpha}_{5}l^{4}R^{ab}%
R^{cd}\mathfrak{R}^{\left(  ef,2\right)  }+6\mathbf{\alpha}_{5}l^{2}%
R^{ab}\mathfrak{R}^{\left(  cd,1\right)  }e^{e}e^{f}\right. \\
&  \left.  +6\mathbf{\alpha}_{5}l^{2}R^{ab}R^{cd}e^{e}h^{\left(  f,1\right)
}+3\mathbf{\alpha}_{5}R^{ab}e^{c}e^{d}e^{e}e^{f}\right)  \delta e^{g}\\
&  +\varepsilon_{abcdefg}\left(  \mathbf{\alpha}_{3}l^{4}R^{ab}R^{cd}%
R^{ef}+3\mathbf{\alpha}_{5}l^{4}R^{ab}R^{cd}\mathfrak{R}^{\left(  ef,1\right)
}+3\mathbf{\alpha}_{5}l^{2}R^{ab}R^{cd}e^{e}e^{f}\right)  \delta h^{\left(
g,1\right)  }\\
&  +\varepsilon_{abcdefg}\left(  \mathbf{\alpha}_{5}l^{4}R^{ab}R^{cd}%
R^{ef}\right)  \delta h^{\left(  g,2\right)  }\\
&  +\varepsilon_{abcdefg}\delta\omega^{ab}\left(  3\mathbf{\alpha}_{1}%
l^{4}R^{cd}R^{ef}T^{g}+3\mathbf{\alpha}_{3}l^{4}R^{cd}R^{ef}\mathfrak{T}%
^{\left(  g,1\right)  }+6\mathbf{\alpha}_{3}l^{4}R^{cd}\mathfrak{R}^{\left(
ef,1\right)  }T^{g}\right. \\
&  \left.  +6\mathbf{\alpha}_{3}l^{2}R^{cd}e^{e}e^{f}T^{g}+3\mathbf{\alpha
}_{5}l^{4}R^{cd}R^{ef}\mathfrak{T}^{\left(  g,2\right)  }+3\mathbf{\alpha}%
_{5}l^{4}\mathfrak{R}^{\left(  cd,1\right)  }\mathfrak{R}^{\left(
ef,1\right)  }T^{g}\right. \\
&  \left.  +2\mathbf{\alpha}_{5}l^{4}k^{\left(  cd,1\right)  }k^{\left(
ef,1\right)  }T^{g}+6\mathbf{\alpha}_{5}l^{4}R^{cd}\mathfrak{R}^{\left(
ef,2\right)  }T^{g}+6\mathbf{\alpha}_{5}l^{4}R^{cd}\mathfrak{R}^{\left(
ef,1\right)  }\left(  \mathfrak{T}^{\left(  g,1\right)  }+k_{\text{ \ }%
h}^{g\text{ }\left(  1\right)  }e^{h}\right)  \right. \\
&  \left.  +6\mathbf{\alpha}_{5}l^{2}\mathfrak{R}^{\left(  cd,1\right)  }%
e^{e}e^{f}T^{g}+3\mathbf{\alpha}_{5}l^{2}R^{cd}e^{e}e^{f}\mathfrak{T}^{\left(
g,1\right)  }+3\mathbf{\alpha}_{5}e^{c}e^{d}e^{e}e^{f}T^{g}\right) \\
&  +\varepsilon_{abcdefg}\delta k^{\left(  ab,1\right)  }\left(
3\mathbf{\alpha}_{3}l^{4}R^{cd}R^{ef}T^{g}+6\mathbf{\alpha}_{5}l^{4}%
R^{cd}\mathfrak{R}^{\left(  ef,1\right)  }T^{g}\right. \\
&  \left.  +3\mathbf{\alpha}_{5}l^{4}R^{cd}R^{ef}\left(  \mathfrak{T}^{\left(
g,1\right)  }+k_{\text{ \ }h}^{g\text{ }\left(  1\right)  }e^{h}\right)
+2\mathbf{\alpha}_{5}l^{2}R^{cd}e^{e}e^{f}T^{g}\right) \\
&  +\varepsilon_{abcdefg}\delta k^{\left(  ab,2\right)  }\left(
3\mathbf{\alpha}_{5}l^{4}R^{cd}R^{ef}T^{g}\right)
\end{align*}
In the event that $\left(  i\right)  $ $\alpha_{1}$ and $\alpha_{3}$ are zero,
$\left(  ii\right)  $ the torsionless condition is imposed and $\left(
iii\right)  $ $k^{\left(  ab,1\right)  }=0,k^{\left(  ab,2\right)
}=0,h^{\left(  a,1\right)  }=0,h^{\left(  a,2\right)  }=0$, is found%

\begin{align}
\delta L_{CS\;\left(  6+1\right)  }^{\mathcal{M}_{7}}  &  =\varepsilon
_{abcdefg}\left(  3\mathbf{\alpha}_{5}R^{ab}e^{c}e^{d}e^{e}e^{f}\right)
\delta e^{g}+\varepsilon_{abcdefg}\left(  3\mathbf{\alpha}_{5}l^{2}%
R^{ab}R^{cd}e^{e}e^{f}\right)  \delta h^{\left(  g,1\right)  }\nonumber\\
&  +\varepsilon_{abcdefg}\left(  \mathbf{\alpha}_{5}l^{4}R^{ab}R^{cd}%
R^{ef}\right)  \delta h^{\left(  g,2\right)  }.
\end{align}

Where we see that in the limit $l\rightarrow0$ we have that the Lagrangian
leads to the Einstein Hilbert term
\begin{equation}
L_{CS\;\left(  6+1\right)  }^{\mathcal{M}_{7}}=\frac{3}{5}\mathbf{\alpha}%
_{5}\varepsilon_{abcdefg}R^{ab}e^{c}e^{d}e^{e}e^{f}e^{g}.
\end{equation}
and the condition $\delta L_{CS\;\left(  6+1\right)  }^{\mathcal{M}_{7}}=0$
leads to the Einstein equations,%

\begin{equation}
\delta L_{CS\;\left(  6+1\right)  }^{\mathcal{M}_{7}}=\varepsilon
_{abcdefg}\left(  3\mathbf{\alpha}_{5}R^{ab}e^{c}e^{d}e^{e}e^{f}\right)
\delta e^{g}+\varepsilon_{abcdefg}\delta\omega^{ab}\left(  3\mathbf{\alpha
}_{5}e^{c}e^{d}e^{e}e^{f}T^{g}\right)  .
\end{equation}

The results show that the $\left(  2p+1\right)  $-dimensional Chern-Simons
actions invariant under the algebra $\mathcal{M}_{2m+1}$ does not always lead
to the action of General Relativity. Indeed, for certain values {}{}of $m$ is
impossible to obtain the Einstein-Hilbert term in the $\left(  2p+1\right)
$-dimensional Chern-Simons Lagrangian invariant under $\mathcal{M}_{2m+1}$.
\ This is because to obtain the term Einstein-Hilbert,\textbf{ }is necessary
the presence of the $\left\langle \mathbf{J}_{a_{1}a_{2}}\mathbf{Z}%
_{a_{3}a_{4}}\mathbf{\cdots Z}_{a_{2p-1}a_{2p}}\mathbf{P}_{2p+1}\right\rangle
$ component of the invariant tensor, which is given by%

\begin{equation}
\left\langle \mathbf{J}_{a_{1}a_{2}}\mathbf{Z}_{a_{3}a_{4}}\mathbf{\cdots
Z}_{a_{2p-1}a_{2p}}\mathbf{P}_{a_{2p+1}}\right\rangle _{\mathcal{M}_{2m+1}%
}\mathbf{=}\left\{
\begin{array}
[c]{c}%
l^{2p-1}\alpha_{2p-1}\left\langle J_{a_{1}a_{2}}\cdots J_{a_{2p-1}a_{2p}%
}P_{a_{2p+1}}\right\rangle _{AdS}\text{, \ \ \ \ \ \ if }m\geq p\text{ \ }\\
0\text{,
\ \ \ \ \ \ \ \ \ \ \ \ \ \ \ \ \ \ \ \ \ \ \ \ \ \ \ \ \ \ \ \ \ \ \ \ \ \ \ \ \ \ \ \ \ \ \ \ \ \ \ \ \ \ \ \ \ \ if
\thinspace\thinspace}m<p.
\end{array}
\right.
\end{equation}

This observation leads to state the following theorem:

\begin{theorem}
Let $\mathcal{M}_{2m+1}$ be the Maxwell type algebra, which is obtained from
$AdS$ algebra by $S_{E}^{\left(  2m-1\right)  }$-expansion resonant reduced$.$
\ If $L_{CS\text{ }(2p+1\mathfrak{)}}^{\mathcal{M}_{2m+1}}$ is a Chern-Simons
Lagrangian $\left(  2p+1\right)  $-dimensional invariant under the
$\mathcal{M}_{2m+1}$-algebra, \ then the $(2p+1)$-dimensional Chern-Simons
Lagrangian, lead to the Einstein-Hilbert Lagrangian in a certain limit of the
coupling constant $l$, if and only if $m\geq p$.
\end{theorem}

The following table shows a set of Chern-Simons Lagrangian $L_{CS\text{
}(2p+1\mathfrak{)}}^{\mathcal{M}_{2m+1}}$, invariant under the Lie algebra
$\mathcal{M}_{2m+1}$, that flow into the General Relativity Lagrangian in a
certain limit:%

\begin{equation}%
\begin{tabular}
[c]{ccccccccc}\cline{1-1}\cline{3-3}%
\multicolumn{1}{|c}{$\mathcal{M}_{3}$} & \multicolumn{1}{|c}{} &
\multicolumn{1}{|c}{$L_{\text{ }CS\text{ }\left(  3\right)  }^{\mathcal{M}%
_{3}}$} & \multicolumn{1}{|c}{} &  &  &  &  & \\\cline{1-1}\cline{3-3}%
\cline{3-4}%
\multicolumn{1}{|c}{$\mathcal{M}_{5}$} & \multicolumn{1}{|c}{} &
\multicolumn{1}{|c}{$L_{\text{ }CS\text{ }\left(  3\right)  }^{\mathcal{M}%
_{5}}$} & \multicolumn{1}{|c}{$L_{\text{ }CS\text{ }\left(  5\right)
}^{\mathcal{M}_{5}}$} & \multicolumn{1}{|c}{} &  &  &  & \\\cline{1-1}%
\cline{3-3}\cline{3-5}%
\multicolumn{1}{|c}{$\mathcal{M}_{7}$} & \multicolumn{1}{|c}{} &
\multicolumn{1}{|c}{$L_{\text{ }CS\text{ }\left(  3\right)  }^{\mathcal{M}%
_{7}}$} & \multicolumn{1}{|c}{$L_{\text{ }CS\text{ }\left(  5\right)
}^{\mathcal{M}_{7}}$} & \multicolumn{1}{|c}{$L_{\text{ }CS\text{ }\left(
7\right)  }^{\mathcal{M}_{7}}$} & \multicolumn{1}{|c}{} &  &  & \\\cline{1-1}%
\cline{3-3}\cline{3-5}%
$\vdots$ &  &  & $\vdots$ &  &  &  &  & \\
$\vdots$ &  &  & $\vdots$ &  &  &  &  & \\\cline{1-1}\cline{3-3}\cline{3-8}%
\multicolumn{1}{|c}{$\mathcal{M}_{2n-1}$} & \multicolumn{1}{|c}{} &
\multicolumn{1}{|c}{$L_{\text{ }CS\text{ }\left(  3\right)  }^{\mathcal{M}%
_{2n-1}}$} & \multicolumn{1}{|c}{$L_{\text{ }CS\text{ }\left(  5\right)
}^{\mathcal{M}_{2n-1}}$} & \multicolumn{1}{|c}{$L_{\text{ }CS\text{ }\left(
7\right)  }^{\mathcal{M}_{2n-1}}$} & \multicolumn{1}{|c}{$\cdots$} &
\multicolumn{1}{|c}{$\cdots$} & \multicolumn{1}{|c}{$L_{\text{ }CS\text{
}\left(  2n-1\right)  }^{\mathcal{M}_{2n-1}}$} & \multicolumn{1}{|c}{}%
\\\cline{1-1}\cline{3-3}\cline{3-9}%
\multicolumn{1}{|c}{$\mathcal{M}_{2n+1}$} & \multicolumn{1}{|c}{} &
\multicolumn{1}{|c}{$L_{\text{ }CS\text{ }\left(  3\right)  }^{\mathcal{M}%
_{2n+1}}$} & \multicolumn{1}{|c}{$L_{\text{ }CS\text{ }\left(  5\right)
}^{\mathcal{M}_{2n+1}}$} & \multicolumn{1}{|c}{$L_{\text{ }CS\text{ }\left(
7\right)  }^{\mathcal{M}_{2n+1}}$} & \multicolumn{1}{|c}{$\cdots$} &
\multicolumn{1}{|c}{$\cdots$} & \multicolumn{1}{|c}{$L_{\text{ }CS\text{
}\left(  2n-1\right)  }^{\mathcal{M}_{2n+1}}$} &
\multicolumn{1}{|c|}{$L_{\text{ }CS\text{ }\left(  2n+1\right)  }%
^{\mathcal{M}_{2n+1}}$}\\\cline{1-1}\cline{3-3}\cline{3-9}%
\end{tabular}
\end{equation}

It is interesting to note that for each dimension $D$ of space-time, we have
the Lagrangian $L_{CS\text{ }\left(  D\right)  }$ invariant under the algebra
$\mathcal{M}_{2n+1}$ contains all other $D$-dimensional Lagrangian valuated in
an algebra $\mathcal{M}_{2m+1}$ with $m<n$. \ So it is always possible to
obtain an action of a lower algebra off the appropriate fields.

\section{\textbf{Born-Infeld Lagrangians invariant under the subalgebra
}$\mathfrak{L}^{\mathcal{M}}$}

In this section is shown that the even-dimensional Einstein-Hilbert Lagrangian
can be obtained from a Born-Infeld Lagrangian in $\left(  2p\right)
$-dimensions invariant under the subalgebra $L^{\mathcal{M}_{2m}}$ of the
algebra $M_{2m}$, if and only if $m\geq p$. However, this is not possible when
$m<p$ for Born-Infeld Lagrangian in $\left(  2p\right)  $-dimensions invariant
under the subalgebra $L^{\mathcal{M}_{2m}}$.

\subsection{\textbf{Born-Infeld lagrangian in }$D=4$\textbf{ invariant under
}$\mathfrak{L}^{\mathcal{M}_{5}}$}

Following the definitions of Ref. \cite{salg2}, let us consider the
$S$-expansion of the Lie algebra $SO\left(  3,1\right)  $ using as a semigroup
the sub-semigroup $S_{0}^{\left(  3\right)  }=\left\{  \lambda_{0},\lambda
_{2},\lambda_{4}\right\}  $\ of semigroup $S_{\mathrm{E}}^{\left(  3\right)
}=\left\{  \lambda_{0},\lambda_{1},\lambda_{2},\lambda_{3},\lambda
_{4}\right\}  .$ After perfoming its $0_{S}(=\lambda_{4})$-reduction, one
finds a new Lie algebra, call it $L^{\mathcal{M}_{5}}$ which is a subalgebra
of the so called $M_{5}$ algebra$,$ whose generators $J_{ab}=\lambda_{0}%
\tilde{J}_{ab},$ $Z_{ab}=\lambda_{2}\tilde{J}_{ab},$ satisfy the commutation
relationships%
\begin{align}
\left[  J_{ab,}J_{cd}\right]   &  =\eta_{cb}J_{ad}-\eta_{ca}J_{bd}+\eta
_{db}J_{ca}-\eta_{da}J_{cb},\nonumber\\
\left[  J_{ab,}Z_{cd}\right]   &  =\eta_{cb}Z_{ad}-\eta_{ca}Z_{bd}+\eta
_{db}Z_{ca}-\eta_{da}Z_{cb,}\\
\left[  Z_{ab,}Z_{cd}\right]   &  =0.\nonumber
\end{align}
In order to write down a Born-Infeld, we start from the two-form
$L^{\mathcal{M}_{5}}$ curvature $F$
\begin{equation}
F=\frac{1}{2}R^{ab}J_{ab}+\frac{1}{2}\left(  D_{\omega}k^{ab}+\frac{1}{l^{2}%
}e^{a}e^{b}\right)  Z_{ab}.
\end{equation}

Using Theorem $VII.2$ of Ref. \cite{salg2}, it is possible to show that the
only non-vanishing components of a invariant tensor for the $\mathfrak{L}%
^{\mathcal{M}_{5}}$ algebra are given by%
\begin{equation}
\left\langle J_{ab}J_{cd}\right\rangle _{\mathfrak{L}^{\mathcal{M}_{5}}%
}=\alpha_{0}l^{2}\varepsilon_{abcd},
\end{equation}%
\begin{equation}
\left\langle J_{ab}Z_{cd}\right\rangle _{\mathfrak{L}^{\mathcal{M}_{5}}%
}=\alpha_{2}l^{2}\varepsilon_{abcd}%
\end{equation}
where $\alpha_{0}$ and $\alpha_{2}$ \ are arbitrary independent constants of
dimensions $\left[  \text{length}\right]  ^{-2}$.

Using the dual procedure of $S$-expansion in terms of the Maurer-Cartan forms
\cite{salg3}, we find that the $4$-dimensional Born-Infeld Lagrangian
invariant under the $\mathfrak{L}^{\mathcal{M}_{5}}$ algebra is given by
\cite{salg1a}%
\begin{equation}
L_{BI\text{ \ }(4)}^{\mathfrak{L}^{\mathcal{M}_{5}}}=\frac{\alpha_{0}}%
{4}\epsilon_{abcd}l^{2}R^{ab}R^{cd}+\frac{\alpha_{2}}{2}\epsilon_{abcd}\left(
R^{ab}e^{c}e^{d}+l^{2}D_{\omega}k^{ab}R^{cd}\right)  . \label{L4BI}%
\end{equation}

Here we can see that the Lagrangian $\left(  \text{\ref{L4BI}}\right)  $ is
split into two independent pieces, one proportional to $\alpha_{0}$ and the
other to $\alpha_{2}$. \ The term proportional to $\alpha_{0}$ corresponds to
the Euler invariant. The piece proportional to $\alpha_{2}$ contains the
Einstein-Hilbert term $\varepsilon_{abcd}R^{ab}e^{c}e^{d}$ plus a boundary
term which contains, besides the usual curvature $R^{ab},$ a bosonic matter
field $k^{ab}$.

Unlike the Born-Infeld Lagrangian the coupling constant $l^{2}$ does not
appear explicitly in the Einstein Hilbert term but accompanies the remaining
elements of the Lagrangian. This allows recover four dimensional the
Einstein-Hilbert Lagrangian in the limit where $l$ equals to zero.

The variation of the Lagrangian, modulo boundary terms, is given by%

\begin{equation}
\delta L_{BI\text{ \ }(4)}^{\mathfrak{L}^{\mathcal{M}_{5}}}=\varepsilon
_{abcd}\left(  \alpha_{2}R^{ab}e^{c}\right)  \delta e^{d}+\varepsilon
_{abcd}\text{ }\delta\omega^{ab}\left(  \alpha_{2}T^{c}e^{d}+\alpha
_{2}k_{\text{ \ }e}^{c}R^{ed}\right)  .
\end{equation}
from which we see that to recover the field equations of general relativity is
not necessary to impose the limit $l=0$. \ $\delta L_{BI\text{ \ }%
(4)}^{\mathfrak{L}^{\mathcal{M}}}=0$ leads to the dynamics of Relativity when
considering the case of a solution without matter ($k^{ab}=0$). This is
possible only in 4 dimensions. However, to recover the field equations of
general relativity in dimensions greater than 4, is necessary to take a limit
of the coupling constant $l$.

\subsection{\textbf{Born-Infeld lagrangian in }$D=4$\textbf{ invariant under
}$\mathfrak{L}^{\mathcal{M}_{7}}$ algebra}

Now, we consider the Born-Infeld lagrangian in $D=4$ invariant under
$\mathfrak{L}^{\mathcal{M}_{7}}$ algebra whose generators satifsy the
following commutation relations
\begin{align}
\left[  J_{ab,}J_{cd}\right]   &  =\eta_{cb}J_{ad}-\eta_{ca}J_{bd}+\eta
_{db}J_{ca}-\eta_{da}J_{cb}\\
\left[  J_{ab,}Z_{cd}^{\left(  1\right)  }\right]   &  =\eta_{cb}%
Z_{ad}^{\left(  1\right)  }-\eta_{ca}Z_{bd}^{\left(  1\right)  }+\eta
_{db}Z_{ca}^{\left(  1\right)  }-\eta_{da}Z_{cb}^{\left(  1\right)  }\\
\left[  J_{ab,}Z_{cd}^{\left(  2\right)  }\right]   &  =\eta_{cb}%
Z_{ad}^{\left(  2\right)  }-\eta_{ca}Z_{bd}^{\left(  2\right)  }+\eta
_{db}Z_{ca}^{\left(  2\right)  }-\eta_{da}Z_{cb}^{\left(  2\right)  }\\
\left[  Z_{ab,}^{\left(  1\right)  }Z_{cd}^{\left(  1\right)  }\right]   &
=\eta_{cb}Z_{ad}^{\left(  2\right)  }-\eta_{ca}Z_{bd}^{\left(  2\right)
}+\eta_{db}Z_{ca}^{\left(  2\right)  }-\eta_{da}Z_{cb}^{\left(  2\right)  }.\\
\left[  Z_{ab,}^{\left(  1\right)  }Z_{cd}^{\left(  2\right)  }\right]   &
=0=\left[  Z_{ab,}^{\left(  2\right)  }Z_{cd}^{\left(  2\right)  }\right]
\end{align}
The two-form curvature $S_{0}^{\left(  3\right)  }$-expanded reduced is%
\begin{align}
F  &  =\frac{1}{2}R^{ab}J_{ab}+\frac{1}{l}T^{a}P_{a}+\frac{1}{2}\left(
D_{\omega}k^{\left(  ab,1\right)  }+\frac{1}{l^{2}}e^{a}e^{b}\right)
Z_{ab}^{\left(  1\right)  }\nonumber\\
&  +\frac{1}{2}\left(  D_{\omega}k^{\left(  ab,2\right)  }+k_{\text{ \ }%
c}^{a\text{ }\left(  1\right)  }k^{cb\left(  1\right)  }+\frac{1}{l^{2}%
}\left[  e^{a}h^{\left(  b,1\right)  }+h^{\left(  a,1\right)  }e^{b}\right]
\right)  Z_{ab}^{\left(  2\right)  }.
\end{align}

Using theorem VII.2 of Ref. \cite{salg2}, it is posible to show that the only
non-vanishing components of a invariant tensor for the $\mathfrak{L}%
^{\mathcal{M}_{7}}$ algebra are given by
\begin{equation}
\left\langle J_{ab}J_{cd}\right\rangle _{\mathfrak{L}^{\mathcal{M}_{7}}%
}=\alpha_{0}\varepsilon_{abcd},
\end{equation}%
\begin{equation}
\left\langle J_{ab}Z_{cd}^{\left(  1\right)  }\right\rangle _{\mathfrak{L}%
^{\mathcal{M}_{7}}}=\alpha_{2}\varepsilon_{abcd},
\end{equation}%
\begin{equation}
\left\langle J_{ab}Z_{cd}^{\left(  2\right)  }\right\rangle _{\mathfrak{L}%
^{\mathcal{M}_{7}}}=\left\langle Z_{ab}^{\left(  1\right)  }Z_{cd}^{\left(
1\right)  }\right\rangle _{\mathfrak{L}^{\mathcal{M}_{7}}}=\alpha
_{4}\varepsilon_{abcd},
\end{equation}
where $\alpha_{0}$, $\alpha_{2}$ and $\alpha_{4}$ \ are arbitrary independent
constants dimensionless.

Using the dual procedure of $S$-expansion in terms of the Maurer-Cartan forms
\cite{salg3}, we find that the $4$-dimensional Born-Infeld Lagrangian
invariant under the $\mathfrak{L}^{\mathcal{M}_{7}}$ algebra is given by
\begin{align}
L_{BI\text{ \ }(4)}^{\mathfrak{L}^{\mathcal{M}_{7}}}  &  =\frac{\alpha_{0}}%
{4}\epsilon_{abcd}R^{ab}R^{cd}+\frac{\alpha_{2}}{2}\epsilon_{abcd}\left(
\mathfrak{R}^{(ab,1)}R^{cd}+\frac{1}{l^{2}}R^{ab}e^{c}e^{d}\right)
\label{L4BI7}\\
&  +\frac{\alpha_{4}}{4}\epsilon_{abcd}\left(  \mathfrak{R}^{(ab,1)}%
\mathfrak{R}^{(cd,1)}+\mathfrak{R}^{(ab,2)}R^{cd}+\frac{2}{l^{2}}%
\mathfrak{R}^{(ab,1)}e^{c}e^{d}+\frac{4}{l^{2}}R^{ab}h^{\left(  c,1\right)
}e^{d}+\frac{1}{l^{4}}e^{a}e^{b}e^{c}e^{d}\right)  ,\nonumber
\end{align}
where%
\begin{align}
\mathfrak{R}^{\left(  ab,1\right)  }  &  =D_{\omega}k^{\left(  ab,1\right)
},\\
\mathfrak{R}^{\left(  ab,2\right)  }  &  =D_{\omega}k^{\left(  ab,2\right)
}+k_{\text{ \ }c}^{a\text{ }\left(  1\right)  }k^{cb\left(  1\right)  }.
\end{align}

The Lagrangian $\left(  \text{\ref{L4BI7}}\right)  $ is split into three
independent pieces, each one proportional to $\alpha_{0}$, $\alpha_{2}$,
$\alpha_{4}$ repectively. \ The term proportional to $\alpha_{0}$ corresponds
to the Euler invariant. The piece proportional to $\alpha_{2}$ contains the
Einstein-Hilbert term $\varepsilon_{abcd}R^{ab}e^{c}e^{d}$ plus a boundary
term which contains, besides the usual curvature $R^{ab},$ a bosonic matter
field $k^{\left(  ab,1\right)  }$.

The variation of the Lagrangian, modulo boundary terms, is given by
\begin{align}
\delta L_{BI\text{ \ }(4)}^{\mathfrak{L}^{\mathcal{M}_{7}}}  &  =\varepsilon
_{abcd}\left(  \frac{\alpha_{2}}{l^{2}}R^{ab}e^{c}+\frac{\alpha_{4}}{l^{2}%
}\mathfrak{R}^{(ab,1)}e^{c}+\frac{\alpha_{4}}{l^{2}}R^{ab}h^{\left(
c,1\right)  }+\frac{\alpha_{4}}{l^{4}}e^{a}e^{b}e^{c}\right)  \delta
e^{d}\nonumber\\
&  +\varepsilon_{abcd}\text{ }\left(  \frac{\alpha_{4}}{l^{2}}R^{ab}%
e^{c}\right)  \delta h^{\left(  d,1\right)  }+\varepsilon_{abcd}\text{ }%
\delta\omega^{ab}\left(  \alpha_{2}k_{\text{ }e}^{c\text{ },\left(  1\right)
}R^{de}+\frac{\alpha_{2}}{l^{2}}T^{c}e^{d}+\frac{\alpha_{4}}{2}k_{\text{ \ }%
e}^{c\text{ \ },\left(  2\right)  }R^{de}\right. \nonumber\\
&  \left.  +\frac{\alpha_{4}}{l^{2}}\left(  D_{\omega}h^{\left(  c,1\right)
}e^{d}-h^{\left(  c,1\right)  }T^{d}\right)  \right)  +\varepsilon
_{acde}\delta\omega^{ab}\left(  \alpha_{2}k_{b}^{\text{ }c,\left(  1\right)
}R^{de}+\alpha_{4}k_{b}^{\text{ }c,\left(  1\right)  }\mathfrak{R}^{\left(
de,1\right)  }\right. \nonumber\\
&  \left.  +\frac{\alpha_{4}}{2}k_{b}^{\text{ }c,\left(  2\right)  }%
R^{de}+\frac{\alpha_{4}}{l^{2}}k_{b}^{\text{ }c,\left(  1\right)  }e^{d}%
e^{e}\right)  +\varepsilon_{abcd}\text{ }\delta k^{\left(  ab,1\right)
}\left(  \frac{\alpha_{4}}{l^{2}}T^{c}e^{d}\right) \nonumber\\
&  +\varepsilon_{acde}\delta k^{\left(  ab,1\right)  }\left(  \alpha_{4}%
\omega_{b}^{\text{ }c}\mathfrak{R}^{\left(  de,1\right)  }+\frac{\alpha_{4}%
}{2}k_{b}^{\text{ }c,\left(  1\right)  }R^{de}\right)  .
\end{align}
where $\mathfrak{T}^{\left(  a,1\right)  }=D_{\omega}h^{\left(  a,1\right)
}+k_{\text{ \ }c}^{a\text{ \ }\left(  1\right)  }e^{c}.$ If we consider the
case where $k^{\left(  ab,1\right)  }=h^{\left(  a,1\right)  }=0\,$, we have%
\begin{align}
\delta L_{BI\text{ \ }(4)}^{\mathfrak{L}^{\mathcal{M}_{7}}}  &  =\varepsilon
_{abcd}\left(  \frac{\alpha_{2}}{l^{2}}R^{ab}e^{c}+\frac{\alpha_{4}}{l^{4}%
}e^{a}e^{b}e^{c}\right)  \delta e^{d}+\varepsilon_{abcd}\text{ }\left(
\frac{\alpha_{4}}{l^{2}}R^{ab}e^{c}\right)  \delta h^{\left(  d,1\right)
}\nonumber\\
&  +\varepsilon_{abcd}\text{ }\delta\omega^{ab}\left(  \frac{\alpha_{2}}%
{l^{2}}T^{c}e^{d}\right)  +\varepsilon_{abcd}\text{ }\delta k^{\left(
ab,1\right)  }\left(  \frac{\alpha_{4}}{l^{2}}T^{c}e^{d}\right)  ,
\end{align}
from where
\begin{align}
\varepsilon_{abcd}R^{ab}e^{c}  &  =0,\\
\varepsilon_{abcd}T^{c}e^{d}  &  =0.
\end{align}
That is, we have obtained the Einstein-Hilbert dynamics in a vacuum without
any restriction on the coupling constant $l$.

\subsection{\textbf{Born-Infeld Lagrangian in }$D=4$\textbf{ invariant under
}$\mathfrak{L}^{\mathcal{M}_{2n+1}}$}

The generators of the $\mathcal{M}_{2n+1}$ algebra satisfy the commutation
relation (\ref{1'}-\ref{3'}). The corresponding $1$-form gauge connection $A$
and the two-form curvature $\mathcal{M}_{2n+1}$ valued $F=dA+A^{2}$ \ are
given in (\ref{1''}) y (\ref{2''}). \ The generators of the $\mathfrak{L}%
^{\mathcal{M}_{2n+1}}$ algebra satisfy the following commutation relation%
\begin{align}
\left[  J_{ab,}J_{cd}\right]   &  =\eta_{cb}J_{ad}-\eta_{ca}J_{bd}+\eta
_{db}J_{ca}-\eta_{da}J_{cb}\nonumber\\
\left[  J_{ab,}Z_{cd}^{\left(  i\right)  }\right]   &  =\eta_{cb}%
Z_{ad}^{\left(  i\right)  }-\eta_{ca}Z_{bd}^{\left(  i\right)  }+\eta
_{db}Z_{ca}^{\left(  i\right)  }-\eta_{da}Z_{cb}^{\left(  i\right)
}\nonumber\\
\left[  Z_{ab,}^{\left(  i\right)  }Z_{cd}^{\left(  j\right)  }\right]   &
=\eta_{cb}Z_{ad}^{\left(  i+j\right)  }-\eta_{ca}Z_{bd}^{\left(  i+j\right)
}+\eta_{db}Z_{ca}^{\left(  i+j\right)  }-\eta_{da}Z_{cb}^{\left(  i+j\right)
},
\end{align}
\ Bearing in mind that the non-zero components tensor invariant, are given by%
\begin{equation}
\left\langle J_{\left(  ab,2i\right)  }J_{\left(  cd,2j\right)  }\right\rangle
=\alpha_{2i+2j}\text{ }\varepsilon_{abcd},
\end{equation}
where $\alpha_{2i+2j}$ are arbitrary independent constants
dimensionless\textbf{ }and where we have defined%
\begin{align*}
J_{ab} &  =\lambda_{0}\tilde{J}_{ab}=J_{\left(  ab,0\right)  }\\
Z_{ab}^{\left(  i\right)  } &  =\lambda_{2i}\tilde{J}_{ab}=J_{\left(
ab,2i\right)  }%
\end{align*}
with\textbf{ }$i=1,...,n-1$.

Using the same procedure used in the previous cases, we found that the
Lagrangian $L_{BI\text{ \ }(4)}^{\mathfrak{L}^{\mathcal{M}_{2n+1}}}$ is given
by
\begin{equation}
L_{BI\text{ \ }(4)}^{\mathfrak{L}^{\mathcal{M}_{2n+1}}}=\frac{\alpha_{2i+2j}%
}{4}\varepsilon_{abcd}F^{\left(  ab,2i\right)  }F^{\left(  cd,2j\right)  }.
\end{equation}

Varying the Lagrangian and considering the case without matter, $\left(
k^{\left(  ab,i\right)  }=h^{\left(  a,j\right)  }=0\,\right)  $, we have%
\begin{align}
&  \delta L_{BI\text{ \ }(4)}^{\mathfrak{L}^{\mathcal{M}_{2n+1}}}%
=\varepsilon_{abcd}\left(  \frac{\alpha_{2}}{l^{2}}R^{ab}e^{c}+\frac
{\alpha_{4}}{l^{4}}e^{a}e^{b}e^{c}\right)  \delta e^{d}+\varepsilon
_{abcd}\text{ }\left(  \frac{\alpha_{i+1}}{l^{2}}R^{ab}e^{c}\right)  \delta
h^{\left(  d,i\right)  }\\
&  +\varepsilon_{abcd}\text{ }\delta\omega^{ab}\left(  \frac{\alpha_{2}}%
{l^{2}}T^{c}e^{d}\right)  +\varepsilon_{abcd}\text{ }\delta k^{\left(
ab,i\right)  }\left(  \frac{\alpha_{i+1}}{l^{2}}T^{c}e^{d}\right)  .
\end{align}
equations leading to the field equations of general relativity
\begin{align}
\varepsilon_{abcd}R^{ab}e^{c}  &  =0,\\
\varepsilon_{abcd}T^{c}e^{d}  &  =0.
\end{align}

\subsection{\textbf{Born-Infeld Lagrangian in }$D=6$\textbf{ invariant under
}$\mathfrak{L}^{\mathcal{M}_{2n}}$ algebra}

It should be noted that the $L^{\mathcal{M}_{2n+1}}$ algebra has the property
of being identical to the $L^{\mathcal{M}_{2n}}$ algebra. However, they have
different origins: The $L^{\mathcal{M}_{2n+1}}$ algebra corresponds to a
reduced $S_{0}^{\left(  2n-1\right)  }$-expansion of the Lorentz algebra, as
we have seen previously, and the $L^{\mathcal{M}_{2n}}$ algebra corresponds to
a reduced $S_{0}^{\left(  2n-2\right)  }$-expansion of the Lorentz algebra,
where the semigroup $S_{0}^{\left(  2n-2\right)  }$ is a subsemigroup of the
semigroup $S_{E}^{\left(  2n-2\right)  }=\left\{  \lambda_{i}\right\}
_{i=0}^{2n-1}$.

It is also interesting to note that the $L^{\mathcal{M}_{2n}}$ algebra can be
used to construct different even-dimensional Born-Infeld type lagrangians.
\ For example, if we consider a reduced $S_{0}^{\left(  4\right)  }$-expansion
of the Lorentz algebra $SO\left(  3,1\right)  $, the $L^{\mathcal{M}_{6}}$
algebra in $D=4$ dimensions is obtained, and if we consider a reduced
$S_{0}^{\left(  4\right)  }$-expansion of the Lorentz algebra $SO\left(
5,1\right)  $ then, we get the $L^{\mathcal{M}_{6}}$ algebra in $D=6$
dimensions. \ In this way, the lagrangians $L_{BI\text{ \ }(4)}^{\mathfrak{L}%
^{\mathcal{M}_{6}}}$ and $L_{BI\text{ \ }(6)}^{\mathfrak{L}^{\mathcal{M}_{6}}%
}$, are invariant under the same algebra $L^{\mathcal{M}_{6}}$, but the
indices in the generators $J_{ab}$, runs over 4 and 6 values, repectively.

These considerations allow the construction of gravitational theories in every
even-dimension. \ However, as discussed below, only in some dimensions it is
possible to obtain General Relativity as a weak coupling constant limit of a
Born-Infeld theory.

\subsubsection{\textbf{Born-Infeld Lagrangian in }$D=6$\textbf{ invariant
under }$\mathfrak{L}^{\mathcal{M}_{4}}$}

The Born-Infeld lagrangian invariant under Lorentz algebra is given by
\begin{equation}
L_{BI}^{(6)}=\frac{\kappa}{6}\epsilon_{abcdef}\left(  R^{ab}R^{cd}R^{ef}%
+\frac{3}{l^{2}}R^{ab}R^{cd}e^{e}e^{f}+\frac{3}{l^{4}}R^{ab}e^{c}e^{d}%
e^{e}e^{f}+\frac{1}{l^{6}}e^{a}e^{b}e^{c}e^{d}e^{e}e^{f}\right)  .
\end{equation}
Following the definitions of Ref. \cite{salg2}, let us consider the
$S$-expansion of the Lie algebra $SO\left(  5,1\right)  $ using $S_{0}%
^{\left(  2\right)  }=\left\{  \lambda_{0},\lambda_{2},\lambda_{3}\right\}  $
as subsemigroup of $S_{E}^{\left(  2\right)  }=\left\{  \lambda_{0}%
,\lambda_{1},\lambda_{2},\lambda_{3}\right\}  $. \ After performing its
$0_{S}$-reduction, one finds the $L^{\mathcal{M}_{4}}$ algebra  which
corresponds to \ a subalgebra of $M_{4}$ algebra. \ The new algebra is
generated by $\left\{  J_{ab},Z_{ab}\right\}  $, where these new generators
can be written as%
\begin{align}
\lambda_{0}\otimes\tilde{J}_{ab} &  =J_{ab},\\
\lambda_{2}\otimes\tilde{J}_{ab} &  =Z_{ab}.
\end{align}
In this case, $\tilde{J}_{ab}$ corresponds to the original generator of
$SO\left(  5,1\right)  \ $and the $\lambda_{\alpha}$ belong to a finite
abelian semigroup $S_{0}^{\left(  2\right)  }$.  Using the invariant tensors
\begin{equation}
\left\langle J_{ab}J_{cd}J_{ef}\right\rangle _{\mathfrak{L}^{\mathcal{M}_{4}}%
}=\frac{4}{3}\alpha_{0}\varepsilon_{abcdef},
\end{equation}%
\begin{equation}
\left\langle J_{ab}J_{cd}Z_{ef}\right\rangle _{\mathfrak{L}^{\mathcal{M}_{4}}%
}=\frac{4}{3}\alpha_{2}\varepsilon_{abcdef},
\end{equation}
we find that the six-dimensional Born-Infeld lagrangian invariant under
$\mathfrak{L}^{\mathcal{M}_{4}}$ algebra is given by
\begin{equation}
L_{BI-(\mathfrak{6)}}^{\mathfrak{L}^{\mathcal{M}_{4}}}=\frac{\alpha_{0}}%
{6}\epsilon_{abcdef}R^{ab}R^{cd}R^{ef}+\frac{\alpha_{2}}{2}\epsilon
_{abcdef}\left(  \mathfrak{R}^{ab}R^{cd}R^{ef}+\frac{1}{l^{2}}R^{ab}%
R^{cd}e^{e}e^{f}\right)
\end{equation}
where $\mathfrak{R}^{ab}=D_{\omega}k^{ab}.$

Note that in this case the $S$-expansion procedure caused the Einstein-Hilbert
term disappeared. This means that in the case of a six-dimensional Born-Infeld
Lagrangian invariant under $\mathfrak{L}^{\mathcal{M}_{4}}$ does not lead to
General Relativity in no limit.

\subsubsection{\textbf{Born-Infeld Lagrangian in }$D=6$\textbf{ invariant
under }$\mathfrak{L}^{\mathcal{M}_{6}}$ \textbf{algebra}}

In this case the 2-form curvature is given by \cite{salg1a}%

\begin{align}
F  &  =\frac{1}{2}R^{ab}J_{ab}+\frac{1}{2}\left(  D_{\omega}k^{\left(
ab,1\right)  }+\frac{1}{l^{2}}e^{a}e^{b}\right)  Z_{ab}^{\left(  1\right)
}\nonumber\\
&  +\frac{1}{2}\left(  D_{\omega}k^{\left(  ab,2\right)  }+k_{\text{ \ }%
c}^{a\text{ }\left(  1\right)  }k^{cb\left(  1\right)  }+\frac{1}{l^{2}%
}\left[  e^{a}h^{\left(  b,1\right)  }+h^{\left(  a,1\right)  }e^{b}\right]
\right)  Z_{ab}^{\left(  2\right)  }.
\end{align}
Using the invariant tensors%

\begin{equation}
\left\langle J_{ab}J_{cd}J_{ef}\right\rangle _{\mathfrak{L}^{\mathcal{M}_{6}}%
}=\frac{4}{3}l^{4}\alpha_{0}\varepsilon_{abcdef},
\end{equation}%
\begin{equation}
\left\langle J_{ab}J_{cd}Z_{ef}\right\rangle _{\mathfrak{L}^{\mathcal{M}_{6}}%
}=\frac{4}{3}l^{4}\alpha_{2}\varepsilon_{abcdef},
\end{equation}%
\[
\left\langle J_{ab}J_{cd}Z_{ef}^{\left(  2\right)  }\right\rangle
_{\mathfrak{L}^{\mathcal{M}_{6}}}=\left\langle J_{ab}Z_{cd}^{\left(  1\right)
}Z_{cd}^{\left(  1\right)  }\right\rangle _{\mathfrak{L}^{\mathcal{M}_{6}}%
}=\frac{4}{3}l^{4}\alpha_{4}\varepsilon_{abcdef},
\]
we find that the six-dimensional Born-Infeld lagrangian invariant under
$\mathfrak{L}^{\mathcal{M}_{6}}$ algebra is given by \
\begin{align*}
L_{BI-(6)}^{\mathfrak{L}^{\mathcal{M}_{6}}}  &  =\frac{\alpha_{0}}{6}%
\epsilon_{abcdef}l^{4}R^{ab}R^{cd}R^{ef}+\frac{\alpha_{2}}{2}\epsilon
_{abcdef}\left(  l^{4}\mathfrak{R}^{\left(  ab,1\right)  }R^{cd}R^{ef}%
+l^{2}R^{ab}R^{cd}e^{e}e^{f}\right) \\
&  +\frac{\alpha_{4}}{2}\epsilon_{abcdef}\left(  l^{4}\mathfrak{R}^{\left(
ab,1\right)  }\mathfrak{R}^{\left(  cd,1\right)  }R^{ef}+l^{4}\mathfrak{R}%
^{\left(  ab,2\right)  }R^{cd}R^{ef}+2Rl^{4ab}R^{cd}h^{\left(  e,1\right)
}e^{f}+l^{2}\mathfrak{R}^{\left(  ab,1\right)  }R^{cd}e^{e}e^{f}\right. \\
&  \left.  +R^{ab}e^{c}e^{d}e^{e}e^{f}\right)
\end{align*}
Varying the Lagrangian and considering the case without matter, $k^{\left(
ab,1\right)  }=h^{\left(  a,1\right)  }=0$, we have%

\begin{align}
\varepsilon_{abcdef}R^{ab}e^{c}e^{d}e^{e}  &  =0,\\
\varepsilon_{abcdef}T^{c}e^{d}e^{e}  &  =0.
\end{align}
which are the Einstein equations in vacuum. Note that if in the Lagrangian
$L_{BI-(6)}^{\mathfrak{L}^{\mathcal{M}_{6}}}$ take the limit $l=0$, we obtain
the Einstein Hilbert term.%
\[
L_{BI-(6)}^{\mathfrak{L}^{\mathcal{M}_{6}}}=\frac{\alpha_{4}}{2}%
\epsilon_{abcdef}R^{ab}e^{c}e^{d}e^{e}e^{f}%
\]

\subsection{\textbf{Born-Infeld Lagrangian in }$D=2n$\textbf{ invariant under
}$\mathfrak{L}^{\mathcal{M}_{2n}}$}

The generators of the algebra $\mathfrak{L}^{\mathcal{M}_{2n}}$ satisfy the
following commutation relations%

\begin{align}
\left[  J_{ab,}J_{cd}\right]   &  =\eta_{cb}J_{ad}-\eta_{ca}J_{bd}+\eta
_{db}J_{ca}-\eta_{da}J_{cb}\nonumber\\
\left[  J_{ab,}Z_{cd}^{\left(  i\right)  }\right]   &  =\eta_{cb}%
Z_{ad}^{\left(  i\right)  }-\eta_{ca}Z_{bd}^{\left(  i\right)  }+\eta
_{db}Z_{ca}^{\left(  i\right)  }-\eta_{da}Z_{cb}^{\left(  i\right)
}\nonumber\\
\left[  Z_{ab,}^{\left(  i\right)  }Z_{cd}^{\left(  j\right)  }\right]   &
=\eta_{cb}Z_{ad}^{\left(  i+j\right)  }-\eta_{ca}Z_{bd}^{\left(  i+j\right)
}+\eta_{db}Z_{ca}^{\left(  i+j\right)  }-\eta_{da}Z_{cb}^{\left(  i+j\right)
},
\end{align}
Theorem VII.2 of Ref. \cite{salg1} allows us to see that the only nonzero
components of the tensor invariant are given by%

\begin{equation}
\left\langle J_{\left(  a_{1}a_{2},i_{1}\right)  }\cdots J_{\left(
a_{2n-1}a_{2n},i_{n}\right)  }\right\rangle =\frac{2^{n-1}l^{2n-2}}{n}%
\alpha_{j}\delta_{i_{1}+\cdots+i_{n}}^{j}\text{ }\varepsilon_{a_{1}\cdots
a_{2n}},
\end{equation}
where $j=0,\cdots,2n-2$ and $\alpha_{j}$ \ are arbitrary independent constants
of dimensions $\left[  \text{length}\right]  ^{2-2n}$.

In this case, the 2-form curvature is given by
\begin{equation}
F=\sum_{k=0}^{n-1}\frac{1}{2}F^{\left(  ab,2k\right)  }J_{\left(
ab,2k\right)  }%
\end{equation}
where%
\begin{equation}
F^{\left(  ab,2k\right)  }=d\omega^{\left(  ab,2k\right)  }+\eta_{cd}%
\omega^{\left(  ac,2i\right)  }\omega^{\left(  db,2j\right)  }\delta_{i+j}%
^{k}+\frac{1}{l^{2}}e^{\left(  a,2i+1\right)  }e^{\left(  b,2j+1\right)
}\delta_{i+j+1}^{k}.
\end{equation}

Using the dual procedure of $S$-expansion in terms of the Maurer-Cartan forms
\cite{salg3}, we find that the $2n$-dimensional Born-Infeld Lagrangian
invariant under the $\mathfrak{L}^{\mathcal{M}_{2n}}$ algebra is given by%

\begin{align}
L_{BI\text{ \ }(2n)}^{\mathfrak{L}_{2n}^{\mathcal{M}}}  &  =\sum_{k=1}%
^{n}l^{2k-2}\frac{1}{2n}\binom{n}{k}\alpha_{j}\delta_{i_{1}+\cdots+i_{n}}%
^{j}\delta_{p_{1}+q_{1}}^{i_{k+1}}\cdots\delta_{p_{n-k}+q_{n-k}}^{i_{n}%
}\nonumber\\
&  \varepsilon_{a_{1}\cdots a_{2n}}R^{\left(  a_{1}a_{2},i_{1}\right)  }\cdots
R^{\left(  a_{2k-1}a_{2k},i_{k}\right)  }e^{\left(  a_{2k+1},p_{1}\right)
}\nonumber\\
&  e^{\left(  a_{2k+2},q_{1}\right)  }\cdots e^{\left(  a_{2n-1}%
,p_{n-k}\right)  }e^{\left(  a_{2n},q_{n-k}\right)  }. \label{lagbi2n'}%
\end{align}
In the $l\rightarrow0$ limit, the only surviving term in $\left(
\text{\ref{lagbi2n'}}\right)  $ is given by $k=1$:%
\begin{align}
\left.  L_{BI\text{ \ }(2n)}^{\mathfrak{L}_{2n}^{\mathcal{M}}}\right\vert
_{l=0}  &  =\frac{1}{2}\alpha_{j}\delta_{i+k_{1}+\cdots+k_{2n-2}}%
^{j}\varepsilon_{a_{1}\cdots a_{2n}}R^{\left(  a_{1}a_{2},i\right)
}e^{\left(  a_{3},k_{1}\right)  }\dots e^{\left(  a_{2n},k_{2n-2}\right)
}\nonumber\\
&  =\frac{1}{2}\alpha_{j}\delta_{2p+2q_{1}+1+\cdots+2_{q_{2n-2}}+1}%
^{j}\varepsilon_{a_{1}\cdots a_{2n}}R^{\left(  a_{1}a_{2},2p\right)
}\nonumber\\
&  e^{\left(  a_{3},2q_{1}+1\right)  }\dots e^{\left(  a_{2n},2_{q_{2n-2}%
}+1\right)  }\nonumber\\
&  =\frac{1}{2}\alpha_{j}\delta_{2\left(  p+q_{1}+\cdots+q_{2n-2}\right)
+2n-2}^{j}\varepsilon_{a_{1}\cdots a_{2n}}R^{\left(  a_{1}a_{2},2p\right)
}\nonumber\\
&  e^{\left(  a_{3},2q_{1}+1\right)  }\dots e^{\left(  a_{2n},2_{q_{2n-2}%
}+1\right)  }.
\end{align}
The only non-vanishing component of this expression occurs for $p=q_{1}%
=\cdots=q_{2n-2}=0$, namely
\begin{align}
\left.  L_{BI\text{ \ }(2n)}^{\mathfrak{L}_{2n}^{\mathcal{M}}}\right\vert
_{l=0}  &  =\frac{1}{2}\alpha_{2n-2}\varepsilon_{a_{1}\cdots a_{2n}}R^{\left(
a_{1}a_{2},0\right)  }e^{\left(  a_{3},1\right)  }\cdots e^{\left(
a_{2n},1\right)  }\nonumber\\
&  =\frac{1}{2}\alpha_{2n-2}\varepsilon_{a_{1}\cdots a_{2n}}R^{a_{1}a_{2}%
}e^{a_{3}}\cdots e^{a_{2n}}.
\end{align}
which is proportional to the Einstein-Hilbert lagrangian.

The results show that the 2p-dimensional Born-Infeld actions invariant under
the algebra $\mathfrak{L}^{\mathcal{M}_{2m}}$ does not always lead to the
action of General Relativity. Indeed, for certain values {}{}of m is
impossible to obtain the Einstein-Hilbert term in the 2p-dimensional
Born-Infeld \textbf{type} Lagrangian invariant under $\mathfrak{L}%
^{\mathcal{M}_{2m}}$. \ This is because to obtain the term Hilbert-Einstein,
is necessary the presence of the $\left\langle J_{a_{1}a_{2}}Z_{a_{3}a_{4}%
}\cdots Z_{a_{2p-1}a_{2p}}\right\rangle $ component of the invariant tensor,
which is given by%

\begin{equation}
\left\langle J_{a_{1}a_{2}}Z_{a_{3}a_{4}}\cdots Z_{a_{2p-1}a_{2p}%
}\right\rangle _{\mathfrak{L}^{\mathcal{M}_{2m}}}=\left\{
\begin{array}
[c]{c}%
l^{2p-2}\alpha_{2p-2}\left\langle J_{a_{1}a_{2}}\cdots J_{a_{2p-1}a_{2p}%
}\right\rangle _{\mathfrak{L}}\text{, \ \ \ \ \ \textbf{\ }if }m\geq p\text{
\ \ \ \ \ \ \ \ \ \ \ \ \ \ \ \ \ \ \ }\\
0\text{,
\ \ \ \ \ \ \ \ \ \ \ \ \ \ \ \ \ \ \ \ \ \ \ \ \ \ \ \ \ \ \ \ \ \ \ \ \ \ \ \ \ \ \ \ \ \ if
\thinspace\thinspace}m<p.\text{ \ \ \ \ \ \ \ \ \ \ \ \ \ \ \ \ }%
\end{array}
\right.
\end{equation}

This observation leads to state the following theorem:

\begin{theorem}
Let $\mathfrak{L}^{\mathcal{M}_{2m}}$ be the algebra obtained from Lorentz
algebra by reduced $S_{0}^{\left(  2m-2\right)  }$-expansion, which
corresponds to a subalgebra of the $\mathcal{M}_{2m}$ algebra $.$ \ If
$L_{BI-2p}^{\mathfrak{L}^{\mathcal{M}_{2m}}}$ is a Born-Infeld type Lagrangian
$\left(  2p\right)  $-dimensional built from the 2-form $\mathfrak{L}%
^{\mathcal{M}_{2m}}$ curvature $F$, \ then the $2p$-dimensional Lagrangian
Born-infeld type, leads to the Lagrangian of General Relativity, in a certain
limit of the coupling constant $l$, if and only if $m\geq p$.
\end{theorem}

The following table shows a set of Born-Infeld type Lagrangian $L_{BI-2p}%
^{\mathfrak{L}^{\mathcal{M}_{2n}}}$, invariant under the Lie algebra
$\mathfrak{L}^{\mathcal{M}_{2n}}$, that flow into the Lagrangian General
Relativity in a certain limit:%

\begin{equation}%
\begin{tabular}
[c]{ccccccccc}\cline{1-1}\cline{3-3}%
\multicolumn{1}{|c}{$\mathfrak{L}^{\mathcal{M}_{4}}$} & \multicolumn{1}{|c}{}
& \multicolumn{1}{|c}{$L_{\text{ }BI\text{ }\left(  4\right)  }^{\mathfrak{L}%
^{\mathcal{M}_{4}}}$} & \multicolumn{1}{|c}{} &  &  &  &  & \\\cline{1-1}%
\cline{3-4}%
\multicolumn{1}{|c}{$\mathfrak{L}^{\mathcal{M}_{6}}$} & \multicolumn{1}{|c}{}
& \multicolumn{1}{|c}{$L_{\text{ }BI\text{ }\left(  4\right)  }^{\mathfrak{L}%
^{\mathcal{M}_{6}}}$} & \multicolumn{1}{|c}{$L_{\text{ }BI\text{ }\left(
6\right)  }^{\mathfrak{L}^{\mathcal{M}_{6}}}$} & \multicolumn{1}{|c}{} &  &  &
& \\\cline{1-1}\cline{3-5}%
\multicolumn{1}{|c}{$\mathfrak{L}^{\mathcal{M}_{8}}$} & \multicolumn{1}{|c}{}
& \multicolumn{1}{|c}{$L_{\text{ }BI\text{ }\left(  4\right)  }^{\mathfrak{L}%
^{\mathcal{M}_{8}}}$} & \multicolumn{1}{|c}{$L_{\text{ }BI\text{ }\left(
6\right)  }^{\mathfrak{L}^{\mathcal{M}_{8}}}$} &
\multicolumn{1}{|c}{$L_{\text{ }BI\text{ }\left(  8\right)  }^{\mathfrak{L}%
^{\mathcal{M}_{8}}}$} & \multicolumn{1}{|c}{} &  &  & \\\cline{1-1}\cline{3-5}%
$\vdots$ &  & $\vdots$ &  &  &  &  &  & \\
$\vdots$ &  & $\vdots$ &  &  &  &  &  & \\\cline{1-1}\cline{3-8}%
\multicolumn{1}{|c}{$\mathfrak{L}^{\mathcal{M}_{2n-2}}$} &
\multicolumn{1}{|c}{} & \multicolumn{1}{|c}{$L_{\text{ }BI\text{ }\left(
4\right)  }^{\mathfrak{L}^{\mathcal{M}_{2n-2}}}$} &
\multicolumn{1}{|c}{$L_{\text{ }BI\text{ }\left(  6\right)  }^{\mathfrak{L}%
^{\mathcal{M}_{2n-2}}}$} & \multicolumn{1}{|c}{$L_{\text{ }BI\text{ }\left(
8\right)  }^{\mathfrak{L}^{\mathcal{M}_{2n-2}}}$} &
\multicolumn{1}{|c}{$\cdots$} & $\cdots$ & \multicolumn{1}{|c}{$L_{\text{
}BI\text{ }\left(  2n-2\right)  }^{\mathfrak{L}^{\mathcal{M}_{2n-2}}}$} &
\multicolumn{1}{|c}{}\\\cline{1-1}\cline{3-9}%
\multicolumn{1}{|c}{$\mathfrak{L}^{\mathcal{M}_{2n}}$} & \multicolumn{1}{|c}{}
& \multicolumn{1}{|c}{$L_{\text{ }BI\text{ }\left(  4\right)  }^{\mathfrak{L}%
^{\mathcal{M}_{2n}}}$} & \multicolumn{1}{|c}{$L_{\text{ }BI\text{ }\left(
6\right)  }^{\mathfrak{L}^{\mathcal{M}_{2n}}}$} &
\multicolumn{1}{|c}{$L_{\text{ }BI\text{ }\left(  8\right)  }^{\mathfrak{L}%
^{\mathcal{M}_{2n}}}$} & \multicolumn{1}{|c}{$\cdots$} & $\cdots$ &
\multicolumn{1}{|c}{$L_{\text{ }BI\text{ }\left(  2n-2\right)  }%
^{\mathfrak{L}^{\mathcal{M}_{2n}}}$} & \multicolumn{1}{|c|}{$L_{\text{
}BI\text{ }\left(  2n\right)  }^{\mathfrak{L}^{\mathcal{M}_{2n}}}$%
}\\\cline{1-1}\cline{3-9}%
\end{tabular}
\end{equation}

It is interesting to note that for each dimension $D$ of space-time, we have
the Lagrangian $L_{BI\text{ }\left(  D\right)  }$ invariant under the
$\mathfrak{L}^{\mathcal{M}_{2n}}$ algebra contains all other $D$-dimensional
Lagrangian evaluated in an $\mathfrak{L}^{\mathcal{M}_{2m}}$ algebra with
$m<n$. \ So it is always possible to obtain an action of a lower algebra off
the appropriate fields.

It is also of interest to note that it was found that, analogously to what
happens in the case of three-dimensiona Chern-Simons gravity, in four
dimensions is not necessary to take the limit $l=0$ to result in General Relativity.

\section{\textbf{Comments and Possible Developments}}

In the present work we have shown that:\ 

$(i)$ standard odd-dimensional General Relativity (without a cosmological
constant) emerges as a weak coupling constant limit of a $(2p+1)$-dimensional
Chern-Simons Lagrangian invariant under $M_{2m+1}$ algebra, if and only if
$m\geq p$. \ 

$(ii)$ when $m<p$, is impossible to obtain odd-dimensional General Relativity
from a $\ (2p+1)$-dimensional Chern-Simons Lagrangian\ invariant under the
$M_{2m+1}$ algebra.

$(iii)$ \ standard even-dimensional General Relativity (without a cosmological
constant) emerges as a weak coupling constant limit of a $\left(  2p\right)
$-dimensional Born-Infeld type Lagrangian invariant under $\mathcal{L}%
^{\mathcal{M}_{2m}}$ algebra, if and only if $m\geq p$.

$(iv)$ \ when $m<p$, is impossible to obtain even-dimensional General
Relativity from a $(2p)$-dimensional Born-Infeld \ type Lagrangian invariant
under $\mathcal{L}^{\mathcal{M}_{2m}}$ algebra.

The toy model and procedure considered here could play an important role in
the context of supergravity in higher dimensions. In fact, it seems likely
that it is possible to recover the standard odd and
even-dimensional\ supergravity from a Chern-Simons and Born-Infeld gravity
theories, in a way very similar to the one shown here. In this way, the
procedure sketched here could provide us with valuable information of what the
underlying geometric structure of Supergravity could be (work in
progress).\bigskip

\begin{acknowledgement}
This work was supported in part by FONDECYT Grants N$^{0}$ 1130653 and by
Universidad de Concepci\'{o}n through DIUC Grant N$^{0}$ 212.011.056-1.0.
Three of the authors (PKC, DM, EKR) were supported by grants from the
Comisi\'{o}n Nacional de Investigaci\'{o}n Cient\'{\i}fica y Tecnol\'{o}gica
CONICYT and from the Universidad de Concepci\'{o}n, Chile.
\end{acknowledgement}

\end{document}